\documentclass[12pt,draftclsnofoot,onecolumn]{IEEEtran}

\usepackage{dsfont}
\usepackage{subcaption}
\usepackage{moreverb}
\usepackage{epsfig}
\usepackage{amsmath,amssymb,amsthm,mathrsfs,amsfonts,dsfont}
\usepackage{adjustbox,lipsum}
\usepackage{algorithm,algorithmic}
\usepackage{amsfonts}
\usepackage{epsfig}
\usepackage{amssymb}
\usepackage{amsmath}
\usepackage{amsthm}
\usepackage{multirow}
\usepackage{rotating}
\usepackage{graphicx}
\usepackage{tabularx}
\usepackage{array}
\usepackage{anyfontsize}
\usepackage{color,soul}
\usepackage{graphicx,dblfloatfix}
\usepackage{epstopdf}
\usepackage{blindtext}
\usepackage{amsmath}
\usepackage{amsthm,amssymb,amsmath,bm}
\usepackage{amsfonts}
\usepackage{epsfig}
\usepackage{amssymb}
\usepackage{amsmath}
\usepackage{cite}
\usepackage{graphicx}
\usepackage{fancyhdr}
\usepackage[font={small}]{caption}
\usepackage{tabularx}
\usepackage{tcolorbox}
\usepackage{cite}
\usepackage{algorithm}
\usepackage{algorithmic}
\usepackage{soul}
\usepackage{setspace}
\usepackage{slashbox}
\usepackage{enumitem}

\DeclareMathOperator*{\argmin}{arg\,min}
\newtheorem{theorem}{Theorem}

\newtheorem{remark}{Remark}
\newtheorem{Pro}{Proposition}

\def\blue{\textcolor{blue}}
\def\red{\textcolor{red}}

\def\black{\textcolor{black}}
\def\brown{\textcolor{brown}}

\allowdisplaybreaks

\graphicspath{{Figure/}}

\usepackage{pifont}
\newcommand{\cmark}{\ding{51}}%
\newcommand{\xmark}{\ding{55}}%

\begin{document}

\title{{AoI Minimization in Status Update Control with Energy Harvesting Sensors}}

\author{Mohammad Hatami\IEEEauthorrefmark{1}, 
Markus~Leinonen\IEEEauthorrefmark{1}, 
and Marian~Codreanu\IEEEauthorrefmark{2}
\thanks{
\IEEEauthorrefmark{1}
Centre for Wireless Communications -- Radio Technologies, University of Oulu, Finland. e-mail: mohammad.hatami@oulu.fi, markus.leinonen@oulu.fi.

\IEEEauthorrefmark{2}
Department of Science and Technology, Link\"{o}ping University, Sweden. e-mail: marian.codreanu@liu.se.


}
}

\maketitle

\vspace{-8mm} 
\begin{spacing}{1.35} 
\begin{abstract}
    Information freshness is crucial for time-critical IoT applications, e.g., monitoring and control. {We consider an IoT status update system with users, energy harvesting sensors, and a cache-enabled edge node.} The users receive time-sensitive information about physical quantities, each measured by a sensor.  {Users demand for the information from the edge node whose cache stores the most recently received measurements from each sensor.} To serve a request, the edge node either commands the sensor to send an update or retrieves the aged measurement from the cache. We aim at finding the best actions of the edge node to minimize {the average AoI of the served measurements at the users, termed \textit{on-demand AoI}.} We model this problem as a Markov decision process and develop reinforcement learning (RL) algorithms: model-based value iteration and model-free Q-learning. We also propose a Q-learning method for the realistic case where the edge node is informed about the sensors’ battery levels only via the status updates. The case under transmission limitations is also addressed. Furthermore, properties of an optimal policy are characterized.  Simulation results show that an optimal policy is a threshold-based policy and that the proposed RL methods significantly reduce the average cost compared to several baselines.

\textbf{Index terms --} Internet of Things (IoT), age of information (AoI), energy harvesting, reinforcement learning (RL), value iteration {algorithm (VIA)}, dynamic programming, Q-learning. 
\end{abstract}
\end{spacing}
\vspace{-2mm} 


\section{Introduction}
Internet of Things (IoT) is an emerging technology to connect different devices {to enable emergent applications} with minimal human intervention.
IoT enables the users to effectively interact with the physical surrounding environment and empower context-aware applications like smart cities \cite{Xu2014IoTSurvey}.
{A typical IoT network consists of multiple wireless sensors which measure physical phenomena and communicate the obtained measurements to a destination for further processing, e.g., to perform distributed target detection \cite{ciuonzo2017generalized}.} Two inherent features of such networks are: 1) stringent energy limitations of battery-powered sensors which, however, may be counteracted by \textit{harvesting} energy\footnote{An alternative approach for ultra-low-power IoT sensors is ambient back-scatter communications; see e.g., \cite{Ciuonzo2019Backscatter,Ji2019Backscatter}.} from environmental sources such as sun, heat, and RF ambient \cite{sudevalayam2010energy,AmbientRFEH2014}, and 2) \textit{transient} nature of data, i.e., the sensors' measurements become outdated after a while.
This calls for the design of IoT sensing techniques where the sensors sample and send a minimal number of measurements to conserve the energy
while providing the end users highly fresh data, as required by time-sensitive applications.

{The freshness} of information  can be quantified by the recently emerged metric, the \textit{age of information} (AoI) {\cite{AoI_Orginal_12,yates19AoI,costa16AoI,sun2019age,Antzela2017age}}.
Formally, AoI is defined as the time elapsed since the latest successfully received status update packet at the destination was generated at a source node. 
{We introduce \textit{on-demand AoI} that represents the AoI at the users restricted to the users' request instants.}
The works that address AoI in IoT networks
can be divided into two main classes: 1) the works that focus on analyzing the AoI in a specific scenario under {their proposed} status update control/scheduling policies \cite{chen2020age, mankar2020throughput, niyato2016novel, pappas2020average, krikidis2019averageRF}, and 2) the works that focus on finding an optimal control/scheduling policy for a specific system. 
For the latter class, there are two main approaches. 
The first approach involves finding an optimal policy by applying different tools from optimization theory \cite{yates2017age, bastopcu2020information,yates2015lazy,wu2017optimal_oneunitenergy,arafa2019Erasure,arafa2019timely,arafa2019age}. Such approaches need exact information about the models and statistics of the environment, e.g., the EH probabilities of sensors.
The second category includes designs relying on dynamic programming and  learning methods \cite{zhu2018caching,leng2019adhoc, zhou2019joint, abd2019reinforcement, bacinoglu2015age_oneunitenergy,tunc2019optimal,leng2019AoIcognitive,Stamatakis2019control,ceran2019RLHARQ}.
In this paper, we focus on this category and find an optimal 
policy that minimizes the AoI about the sensors' measurements received by the users in an EH IoT network.




A particular interest has arisen in designing AoI-aware IoT networks \cite{chen2020age,mankar2020throughput}. 
{In \cite{chen2020age}, a threshold-based age-dependent random access algorithm was proposed for massive IoT networks, in which an IoT device sends an update when its age exceeds a predefined threshold.} 
In \cite{mankar2020throughput}, the authors presented a stochastic geometry analysis for the average AoI in a cellular
IoT network.

{AoI has also been investigated in cache updating systems \cite{yates2017age,bastopcu2020information}. In \cite{yates2017age}, the authors introduced a popularity-weighted AoI metric for updating dynamic content in a local cache, where the content is subjected to version updates.} 
{The authors in \cite{bastopcu2020information} considered a cache updating system with a source, a cache, and a user, and found an analytical expression for the average freshness of the files at the user under the proposed threshold policy.}

The works
\cite{niyato2016novel,krikidis2019averageRF,pappas2020average}
focused on analyzing the AoI in EH IoT networks. The authors in \cite{niyato2016novel} considered a known EH model and proposed a threshold adaptation algorithm to maximize the hit rate in an IoT sensing network. In \cite{pappas2020average}, the authors analyzed the average AoI in a cache enabled status updating system with an EH sensor. In \cite{krikidis2019averageRF}, the author derived a closed-form expression for the average AoI in a wireless powered sensor network.

Age-optimal policies for status update packet transmissions in EH networks have been derived in \cite{yates2015lazy,wu2017optimal_oneunitenergy,arafa2019Erasure,arafa2019timely,arafa2019age} by  using  different  methods  from  optimization theory. {In \cite{yates2015lazy}, the authors derived an optimal policy for an EH source that sends updates to a network interface queue for delivery to a monitoring system.} {In \cite{wu2017optimal_oneunitenergy}, the authors derived age-optimal online policies for an EH sensor having a unit-sized or infinite battery using renewal theory.}
In \cite{arafa2019Erasure}, the authors explored the benefits of erasure status feedback for online timely updating for an EH sensor with a unit-sized battery.
Age-optimal transmission policies for EH two-hop networks  were investigated in \cite{arafa2019timely}. {In \cite{arafa2019age}, the authors derived age-optimal policies for an EH sensor with a finite-sized battery.}

{Several works have developed 
an {AoI-optimal} status update systems by using dynamic programming and learning {based} methods \cite{zhu2018caching,leng2019adhoc, zhou2019joint, abd2019reinforcement, bacinoglu2015age_oneunitenergy,tunc2019optimal,leng2019AoIcognitive,Stamatakis2019control,ceran2019RLHARQ}. 
A commonality in these works is to model the problem as a Markov decision process (MDP), and find
an optimal policy using model-based reinforcement  learning (RL) methods based on dynamic programming, e.g., value iteration algorithm (VIA), and/or model-free RL methods, e.g., Q-learning.} 
A comprehensive survey of RL based methods for autonomous IoT networks was presented in \cite{lei2019deep}. The authors in \cite{zhu2018caching} used deep RL to solve a cache replacement problem with a limited cache size and transient data in an IoT network. Minimizing AoI in a wireless ad hoc network via deep RL was investigated in \cite{leng2019adhoc}.
The authors of \cite{zhou2019joint} derived optimal sampling and updating policies that minimize the average AoI in an IoT monitoring system.
In \cite{abd2019reinforcement}, deep RL was used to minimize AoI in a multi-node monitoring system, in which the sensors are powered through wireless energy transfer {by the destination}.
{The authors of \cite{bacinoglu2015age_oneunitenergy} derived age-optimal sampling instants for an EH sensor with known EH statistics. In \cite{tunc2019optimal}, the authors investigated age-optimal policies where an EH sensor takes advantage of multiple available transmission modes.} In \cite{leng2019AoIcognitive}, the authors studied AoI minimization in cognitive radio EH communications. {In \cite{Stamatakis2019control}, the authors studied age-optimal policies for an EH device that monitors a stochastic process, which can be in either a normal or an alarm state of operation.  In \cite{ceran2019RLHARQ}, the authors studied age-optimal policies for cases where the channel and EH statistics are either known or unknown.}


{Majority of the existing works, including all the above ones, investigate the AoI minimization in cases where the updates are relevant to the monitoring entity at all time moments. Only a few works studied a concept similar to the on-demand AoI herein. In \cite{yin2019only}, the authors introduced the idea of effective AoI (EAoI) 
under a generic request-response model where a server serves
the users with time-sensitive information. They elaborated on the fact that minimizing the time-average EAoI is in general different from minimizing the time-average AoI.
In \cite{Li2021waiting}, the authors studied an information-update system where a user pulls information from servers.
However, in contrast to our paper, the works \cite{yin2019only,Li2021waiting} do not consider energy limitation at the source nodes.}

\subsection{Contributions}
{We consider an IoT status update network that consists of EH IoT sensors, a cache-enabled edge node,
and the users.} 
The users receive time-sensitive information about physical quantities, each of which is measured by a sensor.
{The users demand for the information from the edge node (a gateway) whose cache stores the most recently received measurements of each physical quantity.}
To serve a user's request, the edge node can either command the corresponding sensor to 
send a fresh measurement in the form of status update packet {over an unreliable channel}, or use the
aged data in the cache. {The former enables serving a user with fresh measurement, yet consuming energy from the sensor's battery.} 
The latter prevents the activation of the sensors for every request so that the sensors can utilize the sleep mode to save a considerable amount of energy \cite{niyato2016novel}, but the data received by the users becomes stale. This results in an inherent \textit{trade-off} between the 
AoI at the users and conservation of the sensors' energy in the finite batteries.


{We aim to find the best action of the edge node at each time slot, called an optimal policy, to minimize the average AoI about the physical quantities at the users restricted to the users' request moments, i.e., average {on-demand AoI}.
The on-demand AoI minimization is different from the conventional AoI optimization in that the freshness of information is only important when user(s) need the information.
To tackle this
status update control problem, we derive an MDP model and propose RL based algorithms to obtain optimal policies under different circumstances in the learning environment.}
{To summarize, our main contributions are:} 
\begin{spacing}{1.4}
\begin{itemize}
    \item First, we derive an MDP model for the on-demand AoI minimization problem, calculate the state transition probabilities, and propose a model-based VIA to find an optimal policy.
    \item Then, for the case where the state transition probabilities are unknown, we propose a model-free online Q-learning method to search for an optimal policy. As a practical consideration, we also propose an online method for the 
    realistic
    scenario where the edge node is informed about the sensors' battery levels only via the status updates.
    \item We next derive structural properties of the optimal policy -- obtained by VIA -- 
    and show that the optimal policy has a threshold-based structure with respect to the AoI in a specific scenario.
    \item In addition, we investigate a massive IoT scenario where the edge node can command only a limited number of sensors. In particular, we find an optimal policy and
    propose a low-complexity sub-optimal algorithm.
    \item 
    Extensive numerical experiments are conducted to show that 
    an optimal policy is a threshold-based policy and that the proposed RL algorithms significantly reduce the average on-demand AoI as compared to several baseline policies.
\end{itemize}
\end{spacing}

{Our paper has certain relations to
\cite{wu2017optimal_oneunitenergy,arafa2019Erasure,arafa2019timely,arafa2019age,leng2019adhoc,zhu2018caching,zhou2019joint,abd2019reinforcement,bacinoglu2015age_oneunitenergy,tunc2019optimal,leng2019AoIcognitive,Stamatakis2019control,ceran2019RLHARQ,yin2019only}, yet with the following differences.}
The works \cite{wu2017optimal_oneunitenergy,arafa2019Erasure,arafa2019timely,arafa2019age} focus on a continuous-time single EH sensor and use optimization methods different to the MDP based learning methods herein. 
The works \cite{zhu2018caching, zhou2019joint, leng2019adhoc}, {\cite{yin2019only}}, do not consider energy limitations at the source nodes, whereas we consider EH sensors with finite batteries.
In \cite{abd2019reinforcement}, each time slot is allocated either to one sensor to send an update or to the destination to broadcast RF energy signals to charge the sensors; in our system model, all the users' requests in the network are handled by the edge node at each time slot, and the sensors harvest energy from the environment.
{In \cite{bacinoglu2015age_oneunitenergy,tunc2019optimal,leng2019AoIcognitive,Stamatakis2019control,ceran2019RLHARQ}, the authors studied AoI-optimal policies for a single EH sensor that sends updates to a destination in  cases  where  the  updates are  relevant  to  the  monitoring  entity  at  all  time  moments, whereas we investigate on-demand AoI minimization in IoT networks where
EH sensors {send updates to the users via a cache-enabled edge node.}}
{Different from all the above works, we propose a learning based approach for the case where the edge node is informed about the sensors’ battery levels only via the status update packets, i.e., partial battery knowledge at the edge node.}
{To the best of our knowledge, this is the first work that investigates on-demand AoI in an EH IoT network and proposes MDP based learning approaches for age-aware status update control with 
EH sensors.}
A comparative summary of contributions is presented in Table~\ref{table:comparison}.
Preliminary results of this paper appear in \cite{Hatami-etal-20}.

\begin{table}
\centering
\caption{A comparative summary of contributions of the existing works in contrast to our paper}
\label{table:comparison}
\scalebox{0.7}{
\begin{tabular}{|l|*{15}{c|}}\hline
\backslashbox[40mm]{Feature}{Ref}
&\cite{wu2017optimal_oneunitenergy}&\cite{arafa2019Erasure}&\cite{arafa2019timely}&\cite{arafa2019age}&\cite{zhu2018caching}&\cite{leng2019adhoc}&\cite{zhou2019joint}&\cite{abd2019reinforcement}&\cite{bacinoglu2015age_oneunitenergy}&\cite{tunc2019optimal}&\cite{leng2019AoIcognitive} &\cite{Stamatakis2019control}&\cite{ceran2019RLHARQ}&\cite{yin2019only}& Our\\\hline\hline
On-demand AoI&\xmark&\xmark&\xmark&\xmark&\xmark&\xmark&\xmark&\xmark&\xmark&\xmark&\xmark&\xmark&\xmark&\cmark&\cmark\\\hline
Cache-enabled network controller            &\xmark&\xmark&\xmark&\xmark&\cmark&\xmark&\xmark&\xmark&\xmark&\xmark&\xmark&\xmark&\xmark&\xmark&\cmark\\\hline
Partial battery knowledge&\xmark&\xmark&\xmark&\xmark&\xmark&\xmark&\xmark&\xmark&\xmark&\xmark&\xmark&\xmark&\xmark&\xmark&\cmark\\\hline
Multiple sensors         &\xmark&\xmark&\cmark&\xmark&\cmark&\cmark&\cmark&\cmark&\xmark&\xmark&\xmark&\xmark&\xmark&\xmark&\cmark\\\hline
Multiple users           &\xmark&\xmark&\xmark&\xmark&\cmark&\xmark&\xmark&\xmark&\xmark&\xmark&\xmark&\xmark&\xmark&\cmark&\cmark\\\hline
Energy harvesting        &\cmark&\cmark&\cmark&\cmark&\xmark&\xmark&\xmark&\cmark&\cmark&\cmark&\cmark&\cmark&\cmark&\xmark&\cmark\\\hline
MDP modeling             &\xmark&\xmark&\xmark&\xmark&\cmark&\cmark&\cmark&\cmark&\xmark&\cmark&\cmark&\cmark&\cmark&\cmark&\cmark\\\hline
Unreliable channel       &\xmark&\cmark&\xmark&\xmark&\xmark&\xmark&\cmark&\xmark&\xmark&\cmark&\cmark&\cmark&\cmark&\cmark&\cmark\\\hline
\end{tabular}
}\vspace{-8mm}
\end{table}







\textit{Organization:}
The paper is organized as follows. 
Section \ref{sec_systemmodel} presents the system model and problem definition.
A Markov decision process and definition of optimal policies are presented in Section~\ref{sec_pre_rl}.
Our proposed RL-based status update control algorithms are developed in Section~\ref{sec_rl_algorithms}. {Structural properties of an optimal policy are analytically characterized in Section~\ref{sec_analytic_structure}.}
{The scenario under the transmission limitation is addressed in Section~\ref{sec_limited_bw}.}
Simulation results are presented in Section~\ref{sec_simulation}. 
Concluding remarks are drawn in Section~\ref{sec_conclusions}.

{\textit{Notations}: Vectors and sets are written in boldface lower ($\mathbf{a}$) and calligraphy ($\mathcal{S}$) letters, respectively.
The expectation operation 
is denoted as $\mathbb{E}[\cdot]$. The cardinality of a set $\mathcal{S}$ is denoted as $|\mathcal{S}|$. The indicator function
$\mathds{1}_{\{.\}}$
is equal to $1$ (only) whenever the condition $\{.\}$ is true.}

\section{System Model and Problem Formulation}\label{sec_systemmodel}


\subsection{Network Model}\label{sec_network}
\begin{figure}[t!]
\centering
\includegraphics[width=.45\columnwidth]{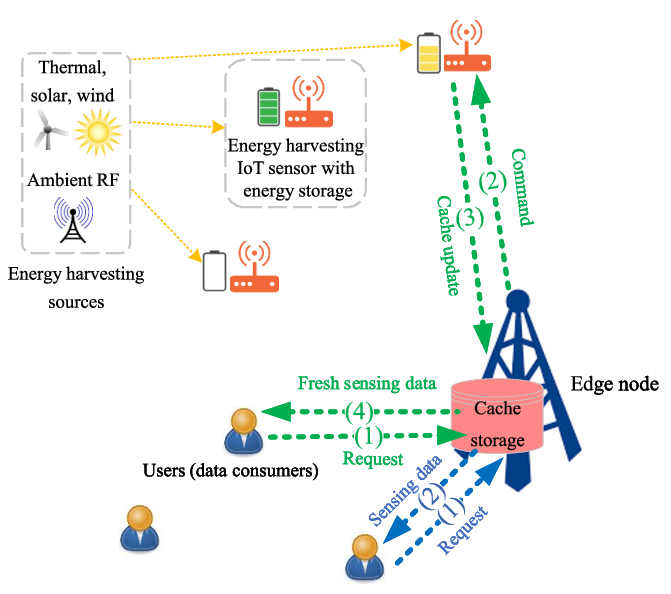}\vspace{-3mm}
\caption{An IoT sensing network consisting of multiple users (\textit{data consumers}), one edge node (i.e., the gateway), and a set of $K$ energy harvesting wireless IoT sensors (\textit{data producers}).  \textcolor{black}{The procedure of serving a request by using fresh data is shown by green lines, whereas the blue lines show the procedure of serving a request by using the previous measurements already existing in the cache.}}
\label{fig_systemmodel}\vspace{-7mm}
\end{figure}

We consider an IoT sensing network consisting of multiple users (\textit{data consumers}), a wireless edge node, and a set {$\mathcal{K}=\left\lbrace 1,\dots,K \right\rbrace$} of $K$ energy harvesting (EH) sensors (\textit{data producers}), as depicted in Fig.~\ref{fig_systemmodel}. Users are interested in time-sensitive information about physical quantities (e.g., temperature or humidity) which are independently measured by the $K$ sensors; formally, sensor $k\in\mathcal{K}$ measures a physical quantity $f_k$. We assume that there is no direct link between the users and the sensors, and the edge node acts as a gateway between them. Thus, the users' requests for the values of $f_k$, $k\in\mathcal{K}$, are served (only) via the edge node.

The system operates in a slotted time fashion, i.e., time is divided into slots labeled with discrete indices ${t \in \mathbb{N}}$. At the beginning of slot $t$, users request for the values of physical quantities $f_k$ from the edge node. Formally, let $r_k(t) \in \{0,1\}$, $t=1,2,\dots$, denote the random process of requesting the value of $f_k$ at the beginning of slot $t$; $r_k(t) = 1$ if the value of $f_k$ is requested and  $r_k(t)=0$ otherwise.
Note that at each time slot, there can be multiple requests arriving at the edge node.

The edge node is equipped with a cache storage that stores the most recently received measurement of each physical quantity $f_k$. Upon receiving a request for the value of $f_k$ at slot $t$ (i.e., $r_k(t)=1$), the edge node can either command sensor $k$ to perform a new measurement and send a \textit{status update}\footnote{{In general, a status  update packet  contains  the  measured  value  of  a  monitored  process  and  a  time  stamp  representing  the time when the sample was generated.}} or use the previous measurement from the local cache, to serve the request.
Let $a_k(t) \in \{0,1\}$ denote the \textit{command} action of the edge node at slot $t$; $a_k(t)=1$ if the edge node commands sensor $k$ to send a status update and $a_k(t)=0$ otherwise. 


{We assume that all the requests that arrive at the beginning of slot $t$ are handled during the same slot $t$. Note that while the communications between the edge node and the users are assumed to be  error-free\footnote{{This assumption is invoked by the fact that the edge node accesses to sufficient power (e.g., a base station connected to a fixed power grid), whereas the sensors rely only on the energy harvested from the environment.
However, it would be straightforward to extend  our proposed approaches to the case where these links are also error-prone.}}, the transmissions from the sensors to the edge node are prone to errors as detailed in Section~\ref{Comm_link}.}

\subsection{Energy Harvesting Sensors}\label{EH_model}
We assume that the sensors rely on the energy harvested from the environment. Sensor $k$ stores the harvested energy into a battery of finite size $B_k$ (units of energy). Formally, let $b_k(t)$ denote the battery level of sensor $k$ at the beginning of slot $t$. Thus, ${b_k(t) \in \{0,\ldots,B_{k}\}}$.

\black{We consider a common assumption  (see e.g., \cite{arafa2019timely,bacinoglu2015age_oneunitenergy,wu2017optimal_oneunitenergy,arafa2019age,michelusi2013transmission}) that transmitting a status update from each sensor to the edge node consumes \textit{one} unit of energy.} Once sensor $k$ is commanded by the edge node (i.e., $a_k(t)=1$), sensor $k$ sends a status update if it has at least one unit of energy in its battery (i.e., $b_k(t) \geq 1$). Let random variable $d_k(t) \in \left\lbrace 0 ,1\right\rbrace$ denote the action of sensor $k$ at slot $t$; $d_k(t)=1$ if sensor $k$ sends a status update to the edge node and $d_k(t)=0$ otherwise.
Accordingly, the relation between the action of sensor $k$ (i.e., $d_k(t)$) and the command action of the edge node (i.e., $a_k(t)$) can be expressed as
\begin{equation}\label{eq_d}
d_k(t) =  a_k(t) \mathds{1}_{\{b_k(t) \geq 1\}},
\end{equation}
Note that quantity $d_k(t)$ in \eqref{eq_d} characterizes also the energy consumption of sensor $k$ at slot $t$.


We model the energy arrivals at the sensors as independent Bernoulli processes with intensities $\lambda_k$, $k\in\mathcal{K}$. 
{This characterizes the discrete nature of the energy arrivals in a slotted-time system, i.e., at each time slot, a sensor either harvests one unit of energy or not (see e.g., \cite{Stamatakis2019control})}.
Let $e_k(t) \in \left\lbrace 0 ,1\right\rbrace $, $t = 1,2,\dots$, denote the \textit{energy arrival process} of sensor $k$. Thus, the probability that sensor $k$ harvests one unit of energy during one time slot is $\lambda_k$, i.e., ${\text{Pr}\{ e_k(t) = 1 \} = \lambda_k}$, $k \in \mathcal{K}$,  $t=1,2,\ldots$.

Finally, using the defined quantities $b_k(t)$, $d_k(t)$, and $e_k(t)$, the evolution of the battery level of sensor $k$ is expressed as
\begin{equation}\label{battery_evo}
b_k(t+1) = \min\left\lbrace  b_k(t)+e_k(t)-d_k(t) , B_k \right\rbrace.
\end{equation}

\subsection{Communication  Between the Edge Node and the Sensors}\label{Comm_link}

\begin{figure}
\centering
\includegraphics[width=.5\columnwidth]{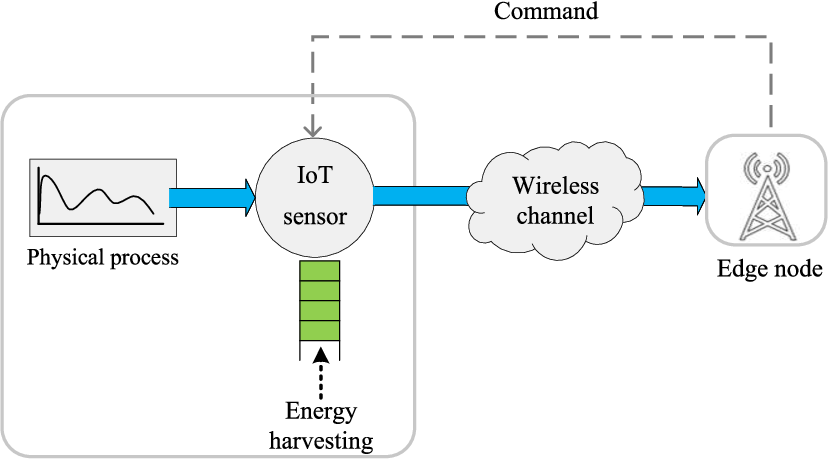}\vspace{-3mm}
\caption{The link between {each sensor} and the edge node consists of an error-free binary command link from the edge node to each sensor and an error-prone wireless communication link from each sensor to the edge node.}\vspace{-7mm}
\label{fig_systemmodel2}
\end{figure}

{We consider} 
an \textit{error-free} binary/single-bit \textit{command} link  from the edge node to each sensor  \cite{ceran2019RLHARQ,arafa2019Erasure}, and an \textit{error-prone} wireless communication link from each sensor  to the edge node, as illustrated in Fig. \ref{fig_systemmodel2}.
If a sensor sends a status update packet to the edge node, the transmission through the wireless link can be either \textit{successful} or \textit{failed}.
{Let $h_k(t) = 1$ denote the  event that a status update from sensor $k$ has been successfully received by the edge node at slot $t$. Otherwise, $h_k(t)=0$ which accounts for both the cases that either 1) sensor $k$ sends a status update but the transmission is failed, or 2) the sensor does not send a status update.}
Let $\xi_k$ be the conditional probability that given that sensor $k$ transmits a status update, it is successfully received by the edge node, i.e.,  ${\text{Pr}\{ h_k(t) = 1 \mid d_k(t) = 1 \} = \xi_k}$, $k \in \mathcal{K}$,  $t=1,2,\ldots$.
Thus, $\xi_k$ represents the \textit{transmit success probability} of the link from sensor $k$ to the edge node.

\subsection{Age of Information}\label{sec_AoI} 
\textit{Age of information} (AoI) is a destination-centric metric that quantifies the freshness of information of a remotely observed random process \cite{AoI_Orginal_12,yates19AoI,costa16AoI}. 
Formally, let  $\Delta_k(t)$ be the AoI
about the physical quantity $f_k$ at the edge node at the beginning of slot  $t$, i.e., the number of time slots elapsed since the generation of the most recently  received status update packet from sensor $k$.
Let $u_k(t)$ denote the most recent time slot in which the edge node received a status update packet from sensor $k$, i.e., $u_k(t) = \max \{t'| t'<t, h_k(t') = 1 \}$; thus, the AoI about $f_k$ can be written as the random process $\Delta_k(t) = t - u_k(t)$.
{We make a common assumption  (see e.g., \cite{zhou2019joint,abd2019reinforcement,leng2019AoIcognitive}) that $\Delta_k(t)$ is upper-bounded by a finite value $\Delta_{k,\text{max}}$, i.e.,  $\Delta_k(t) \in \{1, 2,\ldots ,\Delta_{k,\text{max}}\}$. This is reasonable, because once $\Delta_k(t)$ reaches a high value $\Delta_{k,\text{max}}$, the available measurement about physical process $f_k$ becomes excessively stale/expired, so further counting would be irrelevant.}

{At each time slot, the AoI either drops to one if the edge node receives a status update from the corresponding sensor, or increases by one otherwise. Accordingly, the evolution of  $\Delta_k(t)$ can be written as
\begin{spacing}{1.3}  
\begin{equation}\label{AoI}
\Delta_k(t+1)=
\begin{cases}
1,&\text{if} ~h_k(t)=1, \\
\min \{\Delta_k(t)+1,\Delta_{k,\text{max}}\},&\text{if}~h_k(t)=0,
\end{cases}
\end{equation}
\end{spacing}
which can be expressed compactly as  $\Delta_k(t+1)=\min \Big\{ \big(1-h_k(t) \big) \Delta_k(t)+1,\Delta_{k,\text{max}}\Big\}$.}

\subsection{Cost Function and Problem Formulation}\label{sec_cost}
We consider a cost function that penalizes the  staleness of the requested measurements received by the users. 
{We define the per-sensor immediate cost at slot $t$ as {the \textit{on-demand} AoI as}}
\begin{equation}\label{persensor_cost}
c_k(t) =  r_k(t)  \beta_k \Delta_k(t+1),
\end{equation}
{where $\beta_k \geq 0$ is a pre-defined weight parameter accounting for the importance of the freshness of physical quantity $f_k$,} and $\Delta_k(t+1)$ is the AoI defined in \eqref{AoI}. Note that when the value of $f_k$ is not requested at slot $t$, i.e., $r_k(t) = 0$, the immediate cost becomes $c_k(t) = 0$, as desired.
Moreover, since the requests come at the beginning of slot $t$ and the edge node sends values to the users at the end of the same slot, $\Delta_k(t+1)$ is the effective AoI about $f_k$ seen by the users.

We aim to find  the best action of the edge node at each time slot, i.e., $a_k(t)$, $t = 1,2,\ldots$, $k \in \mathcal{K}$, called an \textit{optimal policy}, that minimizes {the long-term average cost}, 
defined as
\begin{equation}\label{average_cost}
\bar{C}=\displaystyle\underset{T\rightarrow\infty}{\mathrm{lim}}\frac{1}{T}\textstyle\sum_{t=1}^{T}\sum_{k=1}^{K} c_k(t).
\end{equation}

{In order to shed light on the search for such an optimal policy, we next present several points regarding the problem structure.}
{First,} recall {from Section~\ref{sec_network}} that in order to serve the requests for the value of $f_k$ at slot $t$ (i.e., $r_k(t) = 1$), the edge node can either command sensor $k$ to send a status update, i.e., $a_k(t) = 1$, or use the available data in the cache, i.e., $a_k(t) = 0$. The former action (i.e., $a_k(t) = 1$), depending on the battery of sensor $k$ and the situation of the  communication link between sensor $k$ and the edge, \textit{may} lead to having a fresh measurement {(i.e., the AoI drops to one $\Delta_k(t+1) = 1$, minimizing the immediate cost $c_k(t)$ in \eqref{persensor_cost})}, yet at the cost of consuming one unit of energy from the battery of sensor $k$. On the other hand, the latter action  (i.e., $a_k(t) = 0$) provides energy saving at the cost of serving the requests by stale data. 
This introduces an inherent \textit{trade-off} between {(myopically)} minimizing the immediate cost or saving energy for the possible future requests to minimize the cost in a long run.

{It is easy to verify that if there are no requests for the value of $f_k$ at slot $t$ (i.e., $r_k(t) = 0$), the optimal action $a_k(t)$ that minimizes the long-term average cost \eqref{average_cost} is $a_k(t) = 0$. In this case, the immediate cost \eqref{persensor_cost} becomes $c_k(t)=0$, and furthermore, the command action $a_k(t) = 0$ implies $d_k(t)=0$ as per \eqref{eq_d}, leading to energy saving for sensor $k$. Therefore, the search for an optimal policy boils down to finding the optimal actions $a_k(t)$ for the cases with $r_k(t)=1$.}

\begin{remark}\label{rem1}
\normalfont{{For the sake of presentation, we first consider the case where the sensors have independent communication links to the edge node. Accordingly, the edge node can command  any number of sensors at each slot $t$, and these command actions $a_k(t)$, $k\in\mathcal{K}$, are independent across $k$. Thus, the problem of finding the optimal actions $a_k(t)$, $k \in \mathcal{K}$, that minimize \eqref{average_cost} is \textit{separable} across sensors $k \in \mathcal{K}$. Then, in Section~\ref{sec_limited_bw}, we address the case where the edge node can command only a limited number of sensors, which builds on the decoupled case.}}
\end{remark}

{Based on Remark \ref{rem1}, we express the cost in \eqref{average_cost} equivalently as}
{$\bar{C} = \textstyle\sum_{k=1}^{K}\bar{C}_k,$}
where  $\bar{C}_k$ is the average cost associated with sensor $k$, i.e., the \textit{per-sensor} long-term average cost, defined as
\begin{equation}\label{average_cost_persensor}
\bar{C}_k=\displaystyle\underset{T\rightarrow\infty}{\mathrm{lim}}\frac{1}{T}\textstyle\sum_{t=1}^{T} c_k(t),~k = 1, \dots, K.
\end{equation}
Thus, minimizing the system-wise cost in \eqref{average_cost} reduces to minimizing the $K$ per-sensor long-term average costs  in \eqref{average_cost_persensor}. This will be a key factor in developing our reinforcement learning (RL) algorithms in Section~\ref{sec_rl_algorithms}.
{Prior to this, in Section \ref{sec_pre_rl}, we model the considered problem as a Markov decision process (MDP) and {give definitions of} optimal policies, which are needed in our algorithm development.}  






\section{Markov Decision Process and Optimal Policies}\label{sec_pre_rl}
Based on Remark \ref{rem1}, the problem of finding an optimal policy that minimizes the long-term cost in \eqref{average_cost} is separable across the sensors. Thus, we present the derivation of such an optimal policy for a particular sensor $k$ but, clearly, the derivations are valid for any sensor $k\in\mathcal{K}$; the edge node runs in parallel one policy for each sensor in the network.
First, we model the  problem as {an} MDP. Then, we give a formal definition of an optimal policy, followed by introducing the key quantities needed to evaluate and search for such an optimal policy. All these serve as preliminaries for the development of our RL-based algorithms in Section~\ref{sec_rl_algorithms} and Section~\ref{sec_limited_bw}.




\subsection{MDP Modeling}\label{sec_mdp}
\sloppy The MDP model associated with sensor $k$ is defined by the tuple ${\left\lbrace \mathcal{S}_k, \mathcal{A}_k, \mathcal{P}_k \big( s_k(t+1) \big|s_k(t), a_k(t)\big) , c_k\big(s_k(t),a_k(t)\big), \gamma \right\rbrace}$, where
\begin{spacing}{1.4}
\begin{itemize}
\item $\mathcal{S}_k$ is the state set. Let ${s_k(t) \in \mathcal{S}_k}$ denote the state at slot $t$, which is defined as $s_k(t) = \lbrace b_k(t),{\Delta}_k(t) \rbrace $, where 1) $b_k(t)$ is the battery level of sensor $k$ given by \eqref{battery_evo}, i.e., ${b_k(t) \in \{1,2,\ldots,B_k \}}$, and 2) ${\Delta}_k(t)$ is the AoI about the physical quantity $f_k$ in the local cache, i.e., 
${\Delta}_k(t) \in \{1,2,\ldots,\Delta_{k,\text{max}} \}$. 
\item $\mathcal{A}_k = \left\lbrace 0,1 \right\rbrace $ is the action set. The action selected by the edge node  at slot $t$ is denoted by $a_k(t) \in \mathcal{A}_k$ (see Section \ref{sec_network}).
\item $\mathcal{P}_k \big( s_k(t+1) \big|s_k(t), a_k(t)\big)$ is the state transition probability  that maps a state-action pair at  slot $t$ onto a distribution of states at  slot $t + 1$.
\item $c_k(s_k(t),a_k(t))$ is the immediate cost function, i.e., the cost of taking action $a_k(t)$ in state $s_k(t)$, which is also  denoted simply by $c_k(t)$, and is calculated using \eqref{persensor_cost}.
\item $\gamma \in \left[ 0 , 1\right]  $ is a discount factor used to weight the immediate cost relative to the future costs. 
\end{itemize}
\end{spacing}


\subsection{Optimal Policy}
{In an MDP environment, the immediate and long-term costs that the agent -- the edge node -- expects to receive depends on what actions the edge node takes at each time slot, which are selected based on a \textit{policy}.}
Generally, policies can be \textit{stochastic} or \textit{deterministic} \cite[Sect.~1.3]{sutton2018reinforcement}. A stochastic policy $\pi_k = \pi_k(a|s) : \mathcal{S}_k \times \mathcal{A}_k \rightarrow \left[ 0 , 1\right]$ is defined as a mapping from state $s \in \mathcal{S}_k$ to a \textit{probability} of choosing each possible action $a \in \mathcal{A}_k$. 
A deterministic policy is a special case of the stochastic policy where in each state $s \in \mathcal{S}_k$, $\pi_k(a|s) = 1$ for some $a \in \mathcal{A}_k$. {Herein, we use the same notation $\pi_k$ for both stochastic and deterministic policies.}


{The discounted long-term accumulated cost 
is defined as} 
\begin{equation}\label{persensor_accumulated}
C_k(t) = \textstyle\sum_{\tau = 0}^{\infty} \gamma^ 
{\tau} c_k(t+\tau),
\end{equation}
where $c_k(\cdot)$ is the immediate cost calculated using \eqref{persensor_cost}.
Our goal is to find an optimal policy $\pi^*_k$ that minimizes the expected long-term cost in \eqref{persensor_accumulated}, defined as
\begin{equation}\label{opt_policy}
\pi^*_k = \argmin_{\pi_k} \mathbb{E}_{\pi_k}\left[C_k(t) \mid \pi_k \right],
\end{equation}
where $\mathbb{E}_{\pi_k}[\cdot]$ denotes the expected value of $C_k(t)$ given that the edge node follows policy $\pi_k$.

{Having defined an optimal policy, we 
now
present essential definitions as a means to \emph{search} for 
such an optimal policy.} 

\subsection{State-Value and Action-Value Functions}
In order to  evaluate policies and search for an optimal policy $\pi_k^*$, we define the \textit{state-value} and \textit{action-value} functions.  The state-value function specifies how beneficial it is for the edge node to be in a particular state under a policy $\pi_k$.
Formally, the state-value function of state $s \in \mathcal{S}_k$ under a policy $\pi_k$
can be written as
\begin{equation}\label{persensor_value_fcn}
v_{\pi_k} \left( s \right)  \doteq \mathbb{E}_{\pi_k} \left[ C_k(t)|s_k(t) = s \right], ~\forall s\in \mathcal{S}_k.
\end{equation}

The action-value function specifies how beneficial it is for the edge node to perform a particular action in a state under a policy $\pi_k$. Formally, the action-value function
can be written as
\begin{equation}\label{persensor_action_fcn}
q_{\pi_k} \left(s,a \right) \doteq \mathbb{E}_{\pi_k} \left[ C_k(t)|s_k(t) = s, a_k(t) = a \right],~\forall s \in \mathcal{S}_k , a\in \mathcal{A}_k.
\end{equation}

Value functions define a partial ordering over policies. More precisely, a policy $\pi_k$ is defined to be better than or equal to a policy $\pi^\prime_k$ (i.e., $\pi_k \geq \pi^\prime_k$) if and only if $v_{\pi_k}(s) \leq v_{\pi^\prime_k}(s)$
for all $s \in \mathcal{S}_k$ \cite[Sect. 3.6]{sutton2018reinforcement}.
Therefore, an optimal policy $\pi^*_k$ (not necessarily unique), which is better than or equal to all other policies, minimizes the state-value function for all states. 
{Optimal policies achieve the same state-value function (i.e., the \textit{optimal state-value} function) that is defined as}
\begin{equation}\label{opt_v}
v^*_k \left( s \right) \doteq \min_{\pi_k} v_{\pi_k}(s), \forall{s \in \mathcal{S}_k}.
\end{equation}
{The optimal policies also share the same action-value function (i.e., the \textit{optimal action-value} function)  
that is defined as}
\begin{equation}\label{opt_q_def}
q^*_k \left( s,a \right) \doteq \min_{\pi_k} q_{\pi_k} \left( s,a \right), ~ \forall{s \in \mathcal{S}_k}, a \in \mathcal{A}_k. 	
\end{equation}
Accordingly, an optimal {deterministic} policy  $\pi_k^*$ can be obtained by choosing the action $a$ that minimizes $q^*_k \left( s,a \right)$ in each state $s$, which can expressed as
\begin{spacing}{1.3}
\begin{equation}\label{opt_q_1}
\pi^*_k(a|s) = \left\lbrace 
\begin{array}{ll}
{1,} & \mbox{if} ~ a = \argmin_{a \in \mathcal{A}_k} q^*_k(s,a) \\
0, & \mbox{otherwise} \\
\end{array}
\right., \forall s\in \mathcal{S}_k.
\end{equation}
\end{spacing}

According to \eqref{opt_q_1}, the knowledge of the optimal action-value function {$q^*_k(s,a)$} suffices to find an optimal policy $\pi^*_k$.
Also, an optimal policy $\pi^*_k$ can be found via the optimal state-value function $v^*_k(s)$, provided that the state transition probabilities are known.
{In this case, we first find optimal action-value function $q^*_k(s,a)$, given that $v^*_k(s)$ is available for all the states, and then find an optimal policy using \eqref{opt_q_1}.} 
More precisely, under an optimal policy $\pi_k^*$, for  any state $s \in \mathcal{S}_k$ and its possible successor states $s'\in \mathcal{S}_k$, the relationship between the optimal state-value  and action-value functions {can be derived as}
\begin{equation}\label{eq_state_action_relation}
q^*_k \left( s,a \right) = \textstyle\sum_{s' \in \mathcal{S}_k} \mathcal{P}_k \big(s'|s,a\big) \big[ c_k(s,a) + \gamma v_k^*(s')   \big],~\forall s \in \mathcal{S}_k ,~\forall a \in \mathcal{A}_k.
\end{equation}
In summary, one can find an optimal policy if either 1) the optimal action-value function $q^*_k(s,a)$ is available, or 2) the optimal state-value function $v^*_k(s)$ and state transition probabilities $\mathcal{P}_k \big(s'|s,a\big)$ are available. {We next discuss how to find $v^*_k(s)$ and $q^*_k(s,a)$.}

Under $\pi_k^*$, the recursive relationship between the optimal state-value function of state $s$, $v^*_k(s)$, and the optimal state-value function of its possible successor {state $s'$}, $v^*_k(s')$, {is given by}
\begin{equation}\label{bellman_opt_v}
v^*_k \left( s \right) = \min_{a \in \mathcal{A}_k} q^*_k \left( s,a \right) =\min_{a\in\mathcal{A}_k} \textstyle\sum_{s' \in \mathcal{S}_k} \mathcal{P}_k (s'|s,a) \left[ c_k(s,a) + \gamma v^*_k(s') \right],~\forall s \in \mathcal{S}_k.
\end{equation}
The recursive equation in  \eqref{bellman_opt_v} is called the Bellman optimality equation for 
$v_k^*(s)$. 

Assuming the availability of the state transition probabilities $\mathcal{P}_k( s'|s, a)$, 
\eqref{bellman_opt_v} can be used to estimate the optimal state-value function recursively; this is the basis for {our proposed} {VIA} 
in Section \ref{sec_value_alg}. Similar {to \eqref{bellman_opt_v}}, the Bellman optimality equation for 
$q_k^*(s,a)$ is expressed as
\begin{equation}\label{bellman_opt_action}
q^*_k( s,a)=\textstyle\sum_{s' \in \mathcal{S}_k} \mathcal{P}_k (s'|s,a) \left[ c_k(s,a) + \gamma \min_{a' \in \mathcal{A}_k} q_k^*(s',a')   \right],~\forall s \in \mathcal{S}_k, a\in \mathcal{A}_k.
\end{equation}
The Bellman optimality equation in \eqref{bellman_opt_action} is the basis for {our} proposed Q-learning algorithms {devised} in Section~\ref{sec_q_algorithm} {and Section~\ref{sec_rl_partial}}.

\section{Reinforcement Learning Based Status Update Control Algorithms}\label{sec_rl_algorithms}
In this section, we develop three RL-based status update control algorithms for the considered IoT network. The algorithms fall into two main categories: model-based RL and model-free RL. For the MDP model described in Section~\ref{sec_mdp}, we first develop a model-based {VIA} relying on dynamic programming in Section~\ref{sec_value_alg}, 
followed by proposing
a model-free Q-learning algorithm in Section~\ref{sec_q_algorithm}.
As a practical consideration in Section~\ref{sec_rl_partial}, we redefine the state definition of the MDP to propose a Q-learning method for the scenario where the edge node is informed of the sensors' battery levels only via the status update packets.
As a key advantage, the proposed algorithms are simple with low complexity of implementation, which is
important
in practice.



\subsection{Value Iteration Algorithm (VIA)}\label{sec_value_alg}
\textit{Value Iteration} is a model-based RL method that finds the optimal state-value function $v_k^*(s)$, and consequently, an optimal policy $\pi_k^*$ by turning the Bellman optimality equation \eqref{bellman_opt_v} into an iterative update procedure \cite[Section~4.4]{sutton2018reinforcement}.

\subsubsection{Derivation of the State Transition Probabilities}
{In order to apply \eqref{bellman_opt_v}, the {VIA} requires the knowledge of the state transition probabilities of the MDP (see Section \ref{sec_mdp}). These are derived in the following.}
In the considered system model, for a given action $a_k(t)$, the state transition probabilities are functions of both EH rate $\lambda_k$ and transmit success probability $\xi_k$, {which were} defined in Section \ref{EH_model} and \ref{Comm_link}, respectively. The probability of transition from state $s_k(t)$ to state $s_k(t+1)$  under action $a_k(t)$ is given by
\begin{spacing}{1.3} 
\begin{subequations}\label{transiton}
\begin{align}
&\mathcal{P}_k \big(s_k(t+1) \big| s_k(t)=\{ b_k(t) < B_k, {\Delta}_k(t)\} , a_k(t) = 0 \big ) = \notag \\ &\left\lbrace 
\begin{array}{ll}
{\lambda_k,} & s_k(t+1)= \left\lbrace \begin{array}{l}
b_k(t+1) =  b_k(t)+1, \\ {\Delta}_k(t+1)= \min\{{\Delta}_k(t)+1,\Delta_{k,\text{max}}\}
\end{array} \right\rbrace; \\
{1-\lambda_k,} & s_k(t+1)= \left\lbrace  \begin{array}{l}
b_k(t+1) = b_k(t), \\ {\Delta}_k(t+1)= \min\{{\Delta}_k(t)+1,\Delta_{k,\text{max}}\}
\end{array}\right \rbrace; \\
0, & \mbox{otherwise.} \\
\end{array}
\right. \label{transition_case1}\\
&\mathcal{P}_k \big(s_k(t+1) \big| s_k(t)=\{ b_k(t) = B_k, {\Delta}_k(t)\} , a_k(t) = 0 \big ) = 
\notag\\ &{\left\lbrace 
\begin{array}{ll}
{1,} & s_k(t+1)= \left\lbrace \begin{array}{l}
b_k(t+1) = B_k, \\ {\Delta}_k(t+1)= \min\{{\Delta}_k(t)+1,\Delta_{k,\text{max}}\}
\end{array} \right\rbrace; \\
0, & \mbox{otherwise.} \\
\end{array}
\right.} \label{transition_case2}\\
&\mathcal{P}_k \big(s_k(t+1) \big| s_k(t)=\{ b_k(t) = 0, {\Delta}_k(t)\} , a_k(t) = 1 \big ) = \notag \\ &\left\lbrace 
\begin{array}{ll}
{\lambda_k,} & s_k(t+1)= \left\lbrace \begin{array}{l}
b_k(t+1) = 1, \\ {\Delta}_k(t+1)= \min\{{\Delta}_k(t)+1,\Delta_{k,\text{max}}\}
\end{array} \right \rbrace; \\
{1-\lambda_k,} & s_k(t+1)= \left\lbrace  \begin{array}{l}
b_k(t+1) = 0, \\ {\Delta}_k(t+1)= \min\{{\Delta}_k(t)+1,\Delta_{k,\text{max}}\}
\end{array}\right \rbrace; \\
0, & \mbox{otherwise.} \\
\end{array}
\right. \label{transition_case3}\\ 
&\mathcal{P}_k \big(s_k(t+1)\big| s_k(t)=\{  b_k(t) > 0, {\Delta}_k(t)  \},a_k(t) = 1) = \notag \\ &\left\lbrace 
\begin{array}{ll}
{\lambda_k \xi_k}, & s_k(t+1)= \left\lbrace \begin{array}{l}
b_k(t+1) =  b_k(t), \\ {\Delta}_k(t+1)= 1
\end{array}\right \rbrace; \\
\lambda_k (1-\xi_k), & s_k(t+1)= \left\lbrace \begin{array}{l}
b_k(t+1) = b_k(t), \\ {\Delta}_k(t+1)= \min\{{\Delta}_k(t)+1,\Delta_{k,\text{max}}\}
\end{array}\right \rbrace; \\
(1-\lambda_k)\xi_k, & s_k(t+1)= \left\lbrace \begin{array}{l}
b_k(t+1) = b_k(t)-1 , \\ {\Delta}_k(t+1)= 1
\end{array}\right \rbrace; \\
(1-\lambda_k)(1-\xi_k), & s_k(t+1)= \left\lbrace \begin{array}{l}
b_k(t+1) = b_k(t)-1 \\ {\Delta}_k(t+1)=  \min\{{\Delta}_k(t)+1,\Delta_{k,\text{max}}\}
\end{array}\right \rbrace; \\
0 & \mbox{otherwise.} \\
\end{array}
\right. \label{transition_case4} 
\end{align}
\end{subequations}
\end{spacing}
\vspace{1mm}
In brief, the first three expressions \eqref{transition_case1}--\eqref{transition_case3} correspond to cases where sensor $k$ does not send a status update, which leads the AoI about $f_k$ in the local cache to increase by one, whereas in  \eqref{transition_case4} sensor $k$ sends a status update. In \eqref{transition_case4}, four possible events can occur, depending on the success of the transmission attempt and the energy arrivals, characterized by $\xi_k$ and $\lambda_k$, respectively. These cases are detailed in the following.

\begin{spacing}{1.4}  
\begin{itemize}[noitemsep,nolistsep]
\item Case \eqref{transition_case1}: The edge node does not command sensor $k$ (i.e., $a_k(t) = 0$), and thus, the sensor does not send a status update.
\item Case \eqref{transition_case2}: Similar to case \eqref{transition_case1}, but the battery of sensor $k$ is full, and thus, there is no room left for possible harvested energy units.
\item Case \eqref{transition_case3}: Sensor $k$ is commanded, but since its battery is empty (i.e., $b_k(t) = 0$), no update takes place.
\item Case \eqref{transition_case4}: The edge node commands sensor $k$ whose battery is non-empty (i.e., $b_k(t)\ge1$); sensor $k$ sends the status update, consuming one unit of energy. 
\end{itemize}
\end{spacing}

\subsubsection{Algorithm Summary}
{Having defined the state transition probabilities above, we now employ the Bellman optimality equation \eqref{bellman_opt_v} and set up an iterative update procedure, the {VIA}, to find an optimal policy $\pi_k^*$.}
The proposed {VIA}
is presented in Algorithm \ref{value_itr}, which consists of four main stages: 
1) an arbitrary initialization for the optimal state-value function, e.g., $v^*_k(s) = 0$, $\forall s \in \mathcal{S}_k$,
2) in each iteration, update the estimated value for $v^*_k(s)$, $\forall s \in \mathcal{S}_k$, 
3) stop when the maximum difference in $v_k^*(s)$ between two consecutive iterations is below a pre-defined threshold $\theta$, and
4) determine an optimal deterministic policy $\pi^*_k(a|s)$
by using \eqref{eq_state_action_relation} and \eqref{opt_q_1}.


In the {VIA}, it is assumed that the state transition probabilities are known in advance. According to \eqref{transiton}, in order to calculate the state transition probabilities $\mathcal{P}_k(s'|s,a)$, the probabilistic model of the environment, i.e., {EH} probability $\lambda_k$ and the transmit success probability $\xi_k$ need to be known, which are not always available in practice. 
The scenarios under \textit{unknown} state transition probabilities are addressed in the next subsections.


\begin{algorithm}[t]
\caption{Value iteration algorithm (VIA)
}\label{value_itr}
\begin{algorithmic}[1]
\STATE \textbf{Initialize} $v^*_k(s) = 0$, $k \in \mathcal{K}, \forall s \in \mathcal{S}_k$, and determine a small threshold $\theta > 0$.
\FOR{$k = 1, \dots, K$}
\REPEAT [\textit{Update $v^*_k(s)$}]
\STATE $\delta = 0$ \{\textit{For stopping criterion}\} 
\FOR{$s \in \mathcal{S}_k$}
\STATE $\nu = v^*_k(s)$ 
\STATE $v^*_k(s) = \min_{a\in\mathcal{A}_k} \sum_{s' \in \mathcal{S}_k} \mathcal{P}_k (s'|s,a) \left[ c_k(s,a) + \gamma v^*_k(s') \right]$
\STATE $\pi^*_k(a|s) = \mathds{1}_{\{a = \argmin_{a \in \mathcal{A}_k} \sum_{s' \in \mathcal{S}_k} \mathcal{P}_k (s'|s,a) \left[ c_k(s,a) + \gamma v^*_k(s') \right]\}}$
\STATE $\delta = \max \left\lbrace \delta, \left| \nu-v^*_k(s)\right|  \right\rbrace $ \{\textit{Maximum deviation between the iterations}\}
\ENDFOR
\UNTIL{$\delta < \theta$}
\ENDFOR
\STATE {\textbf{Output}: Optimal deterministic per-sensor policies $\pi^*_k(a|s)$, $\forall k \in \mathcal{K}$}
\end{algorithmic}
\end{algorithm}

\subsection{Q-learning Algorithm}\label{sec_q_algorithm}
Q-learning is an \textit{online} model-free RL algorithm that estimates/learns the optimal action-value functions \textit{by experience} and finds an optimal policy iteratively. 
{The main difference to the {VIA} in Section~\ref{sec_value_alg} is that Q-learning does not require the knowledge of the state transition probabilities $\mathcal{P}_k(s'|s,a)$.}

In the Q-learning method, the 
{estimated} action-value function for sensor $k$, denoted as $Q_k(s,a)$, $s \in \mathcal{S}_k$, $a \in \mathcal{A}_k$, directly approximates the optimal action-value function $q^*_k(s,a)$ in \eqref{opt_q_def} \cite[Sect.~6.5]{sutton2018reinforcement}.
%
%
The convergence $Q_k \rightarrow q^*_k$ requires that all state-action pairs continue to be updated. To satisfy this condition, a typical approach is to use the "exploration-exploitation" technique in the action selection. The $\epsilon$-greedy algorithm is one such method that trade-offs exploration and exploitation \cite[Sect.~6.5]{sutton2018reinforcement}.
{Intuitively, exploration is finding more information about the environment, while
exploitation is exploiting known information to minimize the long-term cost.}

Our proposed Q-learning algorithm is presented in Algorithm \ref{q-algorithm}. 
To allow exploration-exploitation, the edge node takes either a random or greedy action at slot $t$; the probability of taking a random action is denoted by $ \epsilon(t) $, and thus, the probability of exploiting the greedy action $a_k(t) = \argmin_{a\in \mathcal{A}_k} Q_k(s_k(t), a)$ is $1-\epsilon(t)$.  Generally, during initial iterations,  it is better to set $\epsilon(t)$ high in order to learn the underlying dynamics, i.e., to allow more exploration. On the other hand, in stationary settings and once enough observations are made, small values of $\epsilon(t)$ become preferable to increase tendency to exploitation.

{As it is shown on line~\ref{line_qupdate_alg} in Algorithm~\ref{q-algorithm}, at each slot/iteration, the value for the Q-function of the current state is updated based on the action taken and the resulting next state, where $\alpha(t)$ represents the learning rate at slot $t$.}


\begin{algorithm}[t!]
\caption{{Online status update control algorithm via  Q-learning}\label{q-algorithm}}
\begin{algorithmic}[1]
\STATE \textbf{Initialize} $Q_k(s,a) = 0$, $\forall s \in \mathcal{S}_k, a \in \mathcal{A}_k$, $k \in \mathcal{K}$
\FOR{each slot $t = 1, 2, 3, \dots$}
\FOR{$k = 1, \dots, K$}
\IF {$r_k(t)=0$} 
\STATE  $a_k(t) = 0$  
\ELSE
\STATE $a_k(t)$ is chosen according to the following probability\\
$ a_k(t)  =\left\{
\begin{array}{ll}
\argmin_{a\in \mathcal{A}_k} Q(s_k(t), a) \hspace{0mm},~\text{w.p.}\,\, \hspace{0mm}  1-\epsilon(t)\\
\textrm{a random action } a\in \mathcal{A}_k \hspace{0mm},~\text{w.p.}\,\, \hspace{0mm} \hspace{5mm} \epsilon(t)
\end{array}
\right.$
\IF{$a_k(t) = 1$} \STATE Command sensor $k$ to send a status update packet
{\STATE \textbf{if} {$b_k(t)>0$} \textbf{then}  $d_k(t) = 1$
\STATE \textbf{else} {$d_k(t) = 0$}}
{\STATE \hspace{-0.6cm} \textbf{else} $d_k(t) = 0$}
\ENDIF
\ENDIF
\STATE Update AoI according to (3) and calculate $c_k(t)$
\ENDFOR
\STATE Wait for the next requests and compute $s_k(t+1)$, $\forall k \in \mathcal{K}$
{\STATE \textbf{for} {$k = 1, \dots, K$} \textbf{do} \{\textit{Update the Q-tables}\}\\
\label{line_qupdate_alg}
$Q_k(s_k(t),a_k(t)) \gets  (1-\alpha(t))Q_k( s_k(t),a_k(t)) + \alpha(t) \big(c_k(t)+\gamma \min_{a\in\mathcal{A}_k} Q_k(s_k(t+1),a)\big)$
\STATE \textbf{end for}}
\ENDFOR
\end{algorithmic}
\end{algorithm}

\subsection{Q-Learning Algorithm with Partial Battery Knowledge}\label{sec_rl_partial}


{In Section \ref{sec_mdp}, we modeled the state of the MDP as ${s_k(t) = \lbrace b_k(t),{\Delta}_k(t) \rbrace}$. Consequently, both the proposed {VIA} in Section~\ref{sec_value_alg} and the Q-learning algorithm in Section~\ref{sec_q_algorithm} rely on the assumption that the edge node knows the \textit{exact} battery levels of the sensors at \emph{each} time slot. This requires continual coordination between the edge node and the sensors, which may not always be feasible. In this section, we consider a realistic environment where the edge node is informed about the battery levels of the sensors only via the \textit{status update packets}. Consequently, the edge node has only \textit{partial} knowledge about the battery levels at each time slot.}


To account for the fact that the edge node is informed about the sensors' battery levels only via the  status update packets, we next modify the state definition of the MDP. {A status update packet} generated at the beginning of slot $t$ consists of the value of physical quantity $f_k$, the battery level of sensor $k$ (i.e., $b_k(t)$), and the timestamp $t$ when the sample was generated. Let $\tilde{b}_k(t)$ denote the \emph{knowledge} about the battery level of sensor $k$ at the edge node at slot $t$. Formally, $\tilde{b}_k(t) = b_k(u_k(t))$, where $u_k(t)$ represents the most recent time slot in which the edge node received a status update packet from sensor $k$, i.e., $u_k(t) = \max \{t'| t'<t, h_k(t') = 1 \}$ (see Section \ref{sec_AoI}). Namely, at time slot $t$, $\tilde{b}_k(t)$ describes what the battery level of sensor $k$ was at the beginning of the most recent time slot at which the edge node received a status update from sensor $k$.  
To stress, the edge node does not know the exact battery level of the sensors at each time slot, but it only has the \emph{partial/outdated} knowledge based on each sensor’s last update.

{Based on the discussions above, we modify the state definition of the MDP defined in Section \ref{sec_mdp} as $s_k(t) = \lbrace \tilde{b}_k(t),{\Delta}_k(t) \rbrace$, thus, the state contains $\tilde{b}_k(t)$ instead of $b_k(t)$. 
However, this state definition makes it impossible to calculate the state transition probabilities and use the {VIA}.
In particular, the underlying decision process is non-Markovian (i.e., not an MDP), caused by the uncertainty that exists in the wireless channel.
For better clarification, consider state $s_k(t) = \lbrace \tilde{b}_k(t),{\Delta}_k(t) \rbrace$ and action $a_k(t) = 0$; the next state is  $s_k(t+1) = \big\{ \tilde{b}_k(t),\min \{\Delta_k(t)+1,\Delta_{k,\text{max}}\} \big\}$ with probability one.  However, given $s_k(t)$ and $a_k(t) = 1$, it is impossible to calculate the state transition probabilities without knowing the actions taken by the edge node during the last $\Delta_k(t) - 1$ slots (i.e., $a_k(t-\Delta_k(t)),\ldots, a_k(t-1)$), implying the non-Markovity in respect to the current state definition. This is because the energy consumed by the sensor is unknown during these $\Delta_k(t) - 1$ slots (in which, by definition, no update has been received); at each such slot, three indistinguishable cases might have happened: 1) the edge node commanded the sensor,  but the transmission was failed, or 2) the edge node commanded the sensor and it could not send a status update because its battery was empty, or 3) the edge node did not command the sensor. While the {first} case consumes one unit of energy from the battery of the sensor, the {second and third cases do not.}
This means that in order to model the underlying decision process as an MDP and be able to calculate the state transition probabilities, the {\textit{exact actions}} taken by the edge node during the last $\Delta_k(t)-1$ slots must be included in the state definition. More precisely, at slot $t$, the state would be defined as $s_k(t) = \big\lbrace \tilde{b}_k(t),{\Delta}_k(t), a_k(t-\Delta_k(t)),\ldots, a_k(t-1) \big\rbrace$.} 
{This, however, makes the state space grow exponentially in terms of $\Delta_k(t)$.}



Despite the aforementioned non-Markovity property of the decision process, we apply the Q-learning presented in Algorithm \ref{q-algorithm} for the partial battery knowledge case with state $s_k(t) = \lbrace \tilde{b}_k(t),{\Delta}_k(t) \rbrace$. Recall that the Q-learning algorithm does not need any prior knowledge about the state transition probabilities. We will assess the performance of this Q-learning method via  simulations in Section~\ref{sec_simulation} and show that it indeed is capable of learning the underlying environment to some extent, thereby significantly outperforming several baseline methods.

\section{Structural Properties of an Optimal Policy}\label{sec_analytic_structure}

\newcommand\Deltabar{\underbar{$\Delta$}}
\newcommand\sbar{\underbar{$s$}}
\newcommand\bbar{\underbar{$b$}}

In this section, we analyze the properties of an optimal policy defined in \eqref{opt_policy}.
We first prove that the optimal state-value function has monotonic properties. Then, we exploit this monotonicity to prove that an optimal policy has a threshold-based structure with respect to the AoI for the case where the link from sensor $k$ to the edge node is error-free, i.e., $\xi_k = 1$. For general cases, threshold-based structures are also numerically illustrated in  Section~\ref{sec_sim_structure}.

Next, we present two propositions that are used to prove properties of an optimal policy expressed in Theorem 1.
\vspace{-2mm}\begin{Pro}\label{prop1}
The optimal state-value function $v_k^*(s)$ is (i) non-decreasing with respect to the AoI, and (ii) non-increasing with respect to the battery level.
\end{Pro}\vspace{-3mm}
The proof is presented in Appendix A.
\vspace{-2mm}\begin{Pro}\label{prop2}
For the case where the link from sensor $k$ to the edge node is perfect, i.e., $\xi_k = 1$, the difference
between the optimal action-value functions for the different actions, denoted by $\delta q_k^*(s) = q_k^*(s,1) - q_k^*(s,0)$, is non-increasing with respect to the AoI.
\end{Pro}\vspace{-3mm}
The proof is presented in Appendix B.
\vspace{-2mm}\begin{theorem}
For the case where the link from sensor $k$ to the edge node is perfect, i.e., $\xi_k = 1$, an optimal policy has a threshold-based structure with respect to the AoI.
\end{theorem}\vspace{-3mm}
\begin{proof}
Proving that an optimal policy has a threshold-based structure with respect to the AoI is equivalent to showing that if the optimal action in state $s = \{b,{\Delta}\}$ is $a_k^*(s) = 1$, then for all the states $\sbar = \{b,\Deltabar\}$, in which $\Deltabar \geq {\Delta}$, the optimal action is $a_k^*(\sbar) = 1$ as well. According to Proposition \ref{prop2}, $q_k^*(\sbar,1) - q_k^*(\sbar,0) \leq q_k^*(s,1) - q_k^*(s,0)$. The optimal action in state $s$ is $a_k^*(s) = 1$, thus $q_k^*(s,1)  - q_k^*(s,0)\leq 0$. Accordingly, $q_k^*(\sbar,1) - q_k^*(\sbar,0) \leq 0$, which shows that the optimal action for state $\sbar$ is $a_k^*(\sbar) = 1$.
\end{proof}

{Besides the fact that analyzing the structures give insight to optimal policies, the inherent threshold-based structure of an optimal policy can be exploited to reduce the computational complexity of the {VIA} (see e.g., \cite{Hsu_modiano2020aoimultiuser_tcm}).} 



\section{Status Update Control under Transmission Limitation} \label{sec_limited_bw}
So far, we assumed that the edge node can command multiple sensors without any constraints at each time slot, which implies the actions $a_k(t)$, $k\in\mathcal{K}$, to be independent 
across $k$.
In this section, we address the case where the edge node can  command only a limited number of sensors. Suppose that, due
to limited radio resources (e.g., bandwidth), the edge node can command no more than $M<K$ sensors at each time slot. Thus, we have the per-slot transmission limitation
\begin{equation}\label{eq_st_hard_bw}
    \textstyle\sum_{k = 1}^{K} a_k(t) \leq M,~\forall t.
\end{equation}
The constraint \eqref{eq_st_hard_bw} \textit{couples} the actions $a_k(t)$, $k\in\mathcal{K}$, and thus,
finding an optimal policy under the transmission constraint is \textit{not separable} across the sensors.



We next model the problem of finding an optimal policy under the transmission constraint \eqref{eq_st_hard_bw} as an MDP. By defining the state similarly as in the per-sensor MDP of Section~\ref{sec_mdp} while incorporating the coupling constraint into the action set allows us to use the developed RL methods of Section~\ref{sec_rl_algorithms}. 
{Due to the coupling constraint, the complexity of the solution grows exponentially by increasing the number of sensors $K$. Thus, as a practical consideration, we also propose a sub-optimal algorithm for which the complexity increases only linearly in $K$. The performance of the proposed sub-optimal solution is numerically demonstrated to be close to the optimal solution in Section~\ref{subsec:sim_coupled}.}



\subsection{MDP Modeling} The problem of finding an optimal policy under  the transmission constraint is modeled as an MDP, defined by the tuple ${\{\mathcal{S},\mathcal{A},\mathcal{P}(\mathbf{s}(t+1)|\mathbf{s}(t),\mathbf{a}(t)), c(\mathbf{s}(t),\mathbf{a}(t))\}}$, where
\begin{spacing}{1.4}
\begin{itemize}
    \item The state set $\mathcal{S}$ is defined as $\mathcal{S} = \mathcal{S}_1 \times \cdots \times \mathcal{S}_K$; the state space dimension is ${|\mathcal{S}| = \prod_{k = 1}^K (B_k+1)\Delta_{k,\textrm{max}}}$. The state of the system at slot $t$ is defined as ${\mathbf{s}(t) = \big( s_1(t),\ldots,s_K(t)\big) \doteq \big(s_k(t)\big)_{k = 1}^{K}}$, {where $s_k(t)$ is defined in  Section~\ref{sec_mdp}}.
    \item  The action set $\mathcal{A}$ is defined as $\mathcal{A}=\big\{(a_1,\ldots,a_K) \mid a_k\in\mathcal{A}_k = \{0,1\},\;\sum_{k=1}^{K}a_k\le{M}\big\}$; the action space dimension is $|\mathcal{A}| = \sum_{m = 0}^M \binom{K}{m}$. The action selected by the edge node at slot $t$ is denoted by $\mathbf{a}(t) = \big(a_k(t)\big)_{k = 1}^{K}$, {where $a_k(t)$ is defined in  Section~\ref{sec_mdp}}.
    \item The state transition probability $\mathcal{P}(\mathbf{s}(t+1)|\mathbf{s}(t),\mathbf{a}(t))$ is calculated as
    \begin{equation}
        \mathcal{P} \big(\mathbf{s}(t+1) \big| \mathbf{s}(t), \mathbf{a}(t) \big ) = \textstyle\prod_{k = 1}^{K} \mathcal{P}_k \big(s_k(t+1) \big| s_k(t), a_k(t) \big ),
    \end{equation}
    {where $\mathcal{P}_k \big(s_k(t+1) \big| s_k(t), a_k(t) \big )$ is calculated according to \eqref{transition_case1}-\eqref{transition_case4}.}
    \item The immediate cost function $c(\mathbf{s}(t),\mathbf{a}(t))$, denoted simply by $c(t)$, is calculated as 
    ${c(t) = \sum_{k = 1}^{K} c_k(t)}$, {where $c_k(.)$ is defined in  Section~\ref{sec_mdp}}.
\end{itemize}
\end{spacing}

\subsection{Optimal and Sub-optimal Algorithms}
\subsubsection{Optimal Policy} An optimal policy under the transmission constraint can be found by following the steps in Section~\ref{sec_rl_algorithms} and using the developed learning methods, i.e., {VIA} or Q-learning.
Because the state and action spaces grow exponentially with respect to the number of sensors, finding an optimal policy is tractable
only for a small number of sensors.
{More precisely, finding an optimal policy is PSPACE-hard which is similar to NP-hard except that the space (i.e., the size of computer memory) is the main limiting factor
\cite{Papadimitriou-PSPASEhard-1999} \cite[Chap.~6]{gittins2011multi}.} {The structural properties of the optimal policy -- obtained by VIA -- can be obtained by following the same steps as in Section~\ref{sec_analytic_structure}, but due to the space limitation, we omit it.}


\subsubsection{Sub-optimal Policy}
In order to reduce the {exponential complexity due to the coupling constraint \eqref{eq_st_hard_bw}} and deal with practical {massive IoT} scenarios, we propose the following sub-optimal policy. First, we \textit{ignore} the constraint \eqref{eq_st_hard_bw}, and find the optimal per-sensor policies $\pi_k^\star$, $k \in \mathcal{K}$, as discussed in Section~\ref{sec_rl_algorithms}, either by using {VIA} or Q-learning.
Then, we \textit{truncate} the scheduling policy to satisfy the constraint \eqref{eq_st_hard_bw} as follows.
At slot $t$, let $\mathcal{X}(t) = \{k\mid a_k(t) = 1, k \in \mathcal{K}\} \subseteq \mathcal{K}$ denote the set of sensors that are commanded under the optimal per-sensor policies $\pi_k^\star$, $k \in \mathcal{K}$. The truncation step separates into two cases: 1) if $|\mathcal{X}(t)| \leq M$, the edge node {simply} commands all of the sensors in $\mathcal{X}(t)$, and 2) otherwise, the edge node commands only the $M$ sensors from $\mathcal{X}(t)$ that have \textit{the largest AoI}. In this regard, the truncation policy conforms to a myopic strategy in that it prioritizes updating the sensors with the highest AoI to minimize the immediate cost. 

\begin{remark}
\normalfont{
{For the case with no energy limitations at the source nodes, 
a Whittle index policy can be obtained which is asymptotically optimal and has low complexity. For instance, in \cite{Hsu_modiano2020aoimultiuser_tcm}, scheduling multiple sensors with a transmission constraint was modeled as a restless multi-armed bandit (RMAB) and a Whittle index policy was obtained. In RMAB, at each time slot, a specific subset
of ``arms''
is selected by the decision maker \cite[Chap.~6]{gittins2011multi}.
In order to cast our problem as an RMAB and be able to find a Whittle index policy, we first need to ensure that, for an optimal policy, \textit{exactly} $M$ sensors are commanded by the edge node at \textit{each} time slot. However, it is clear that in our system model, commanding exactly $M$ sensors at each time slot is highly sub-optimal. This is because of the energy harvesting nature of the sensors. Namely, when the battery level (or the AoI) is low, it is optimal \textit{not} to command the sensor. 
Inspired by the procedure of finding a Whittle index policy \cite[Chap.~6]{gittins2011multi}, we could start by relaxing the per-slot transmission constraint to the long-term average constraint, and decouple the problem along the sensors by using the Lagrange function. Then, by applying the constrained MDP (CMDP) concepts, we can find an optimal policy for the relaxed decoupled problem. Here, there are two main challenges: 1) properly modifying the optimal relaxed policy to satisfy the per-slot constraint, and 2) mathematical analysis to show the above policy is asymptotically optimal. Studying these aspects will be striven for in our future work.}
}
\end{remark}


\section{Simulation Results}\label{sec_simulation}
In this section, we numerically analyze the structural properties of an optimal policy obtained by the VIA. Moreover,
simulation results are presented to demonstrate the performance of the proposed VIA summarized in Algorithm~\ref{value_itr}, the proposed Q-learning algorithms -- Q-learning with exact and partial battery knowledge -- obtained by Algorithm \ref{q-algorithm}, and the proposed algorithms  under the transmission limitations -- optimal and sub-optimal -- developed in Section~\ref{sec_limited_bw}.



\subsection{Simulation Setup}
{The simulation setup is as the following, unless otherwise stated.
We consider $K=3$ EH sensors, i.e.,  $\mathcal{K}=\left\lbrace 1,2,3 \right\rbrace$.
Each sensor $k \in \mathcal{K}$ has a battery with capacity $B_k = 15$ units of energy.} 
At each time slot, the probability that the value of $f_k$ is requested (i.e., $r_k(t) = 1$)  is denoted by $p_k$, i.e., $\mathrm{Pr}\{r_k(t) = 1\} = p_k$.
We set $p_k=0.15$, $k\in\mathcal{K}$.
For the {VIA}, we set {the threshold parameter as} $\theta = 0.001$. 
For the Q-learning method, we set $\epsilon(t)  =0.02+ 0.98 e ^{-\epsilon_\textrm{d}t}$ with decay parameter $\epsilon_\textrm{d} = 10^{-7}$. The learning rate is set to $\alpha(t) = 0.5$ during the first $1/\epsilon_\textrm{d}=10^7$ slots and after that $\alpha(t) = 0.01$. {Table~\ref{tab:sim_param} summarizes the default simulation parameters.}



\begin{table}[H]
\centering
\caption{\label{tab:sim_param}{Default simulation parameters}}
\scalebox{0.8}{
\begin{tabular}{|c |c || c |c|}
 \hline
 Parameter                          & Value& Parameter                         & Value       \\ [0.5ex]\hline\hline
 Number of sensors ($K$)            & $3$  & Discount  factor ($\gamma$)                & $0.99$     \\ \hline
 Capacity of the batteries ($B_k$)  & $15$ & Maximum deviation error in VIA ($\theta$)  & $0.001$    \\\hline
 The  weight  parameters ($\beta_k$)& $1.0$& AoI upper-bound ($\Delta_{k,\textrm{max}}$)& $127$  \\[0.5ex] \hline
\end{tabular}
}
\end{table}


\subsection{Structure of an Optimal Deterministic Policy}\label{sec_sim_structure}
We analyze the structural properties of an optimal deterministic policy obtained by {the} {VIA} for a particular sensor, e.g., sensor 1, and investigate the effect of the {EH} probability $\lambda_1$ and transmit success probability $\xi_1$. 

{Fig.~\ref{fig_lambda}} illustrates the structure of the obtained optimal deterministic policy for different values of the {EH} probability $\lambda_1$ with the transmit success probability $\xi_1 = 0.9$. Each point represents a potential state of the system {as a pair of values of the battery level and AoI, $(b,{\Delta})$}. {In particular,} a red circle indicates that the optimal action {in a given state} is that the edge node does not command the sensor (i.e., $a = 0$), and a blue square  indicates that the optimal action is that the edge node commands the sensor (i.e., $a = 1$). {The set of blue points
is referred to as the \textit{command region} hereinafter.}

From Fig.~\ref{fig_lambda}(a)--(d), we observe that the optimal deterministic policy has a \textit{threshold-based} structure with respect to the battery level and the AoI, which can be expressed as follows:
\begin{spacing}{1.3}
\begin{enumerate}
\item If the optimal action in state $s = \{b,{\Delta}\}$ is $a = 1$, then for all the states $s^\prime = \{b^\prime,{\Delta}\}$, in which $b^\prime \geq b$, the optimal action is $a = 1$ as well.
\item If the optimal action in state $s = \{b,{\Delta}\}$ is $a = 1$, then for all the states $s^\prime = \{b,{\Delta}^\prime\}$, in which ${\Delta}^\prime \geq {\Delta}$, the optimal action is $a = 1$ as well\footnote{In Section~\ref{sec_analytic_structure}, we analytically proved this statement for the special case $\xi_k = 1$. In this section, the numerical results show that an optimal policy has a threshold-based structure with respect to the AoI for all the values of $\xi_k$ as well.}.
\end{enumerate}
\end{spacing}
To exemplify this threshold-based structure in Fig.~\ref{fig_lambda}(a), consider point
$(5, 17)$. Since the optimal action at the point
$(5, 17)$ is $a=1$, we observe that the optimal action at {all} the points $(5, {\Delta})$ where ${\Delta} \geq 17$, and all the points $(b,17)$ where $b \geq 5$, is also $a = 1$.

By comparing Figs.~\ref{fig_lambda}(a)--(d) with each other, we observe that the command region (i.e., the set of blue square points) enlarges by increasing the {EH} probability $\lambda_1$. This is due to the fact that since the sensor harvests energy more often, the edge node commands the sensor to send fresh measurements more often. Note that Fig.~\ref{fig_lambda}(d) is associated with an extreme case in which the edge node always harvests energy at each time slot; in this case, there is always at least one unit of energy available in the battery of the sensor, and thus, for all the states  with $b\geq1$, the optimal action is $a = 1$.

\newcommand\fw{.35}
\newcommand\ft{Lambda}
\begin{figure*}[t]
\centering
\subfloat [$\lambda_1 = 0.005$]{%
\includegraphics[width=\fw \columnwidth]{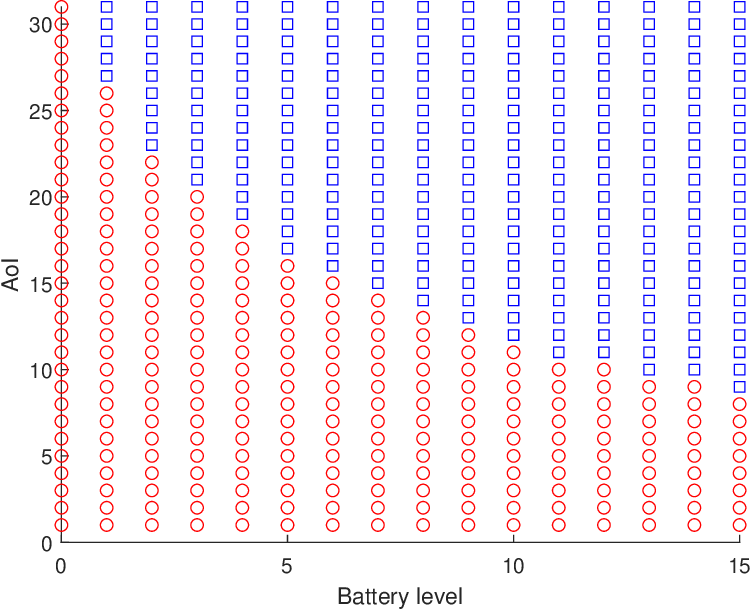}%
\label{fig1_lambda}%
}\qquad
\subfloat [$\lambda_1 = 0.04$]{%
\includegraphics[width=\fw \columnwidth]{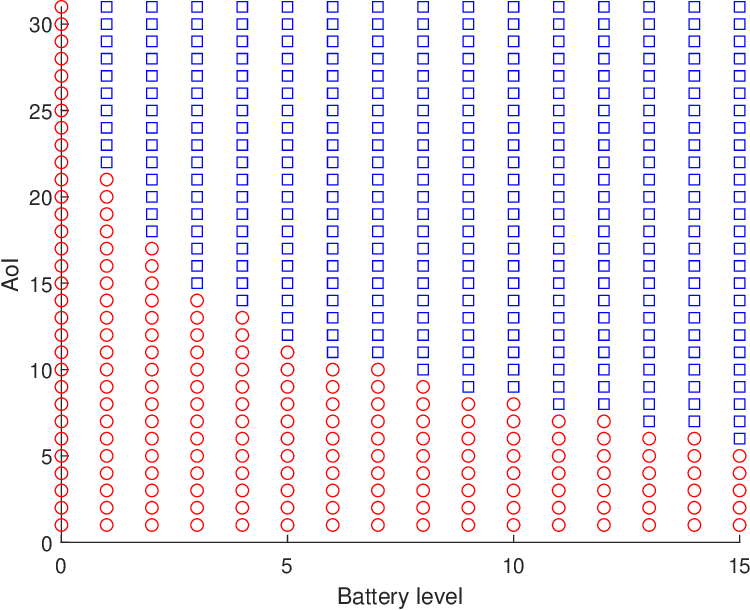}%
\label{fig2_lambda}%
}\\
\subfloat[$\lambda_1 = 0.08$]{%
\includegraphics[width=\fw \columnwidth]{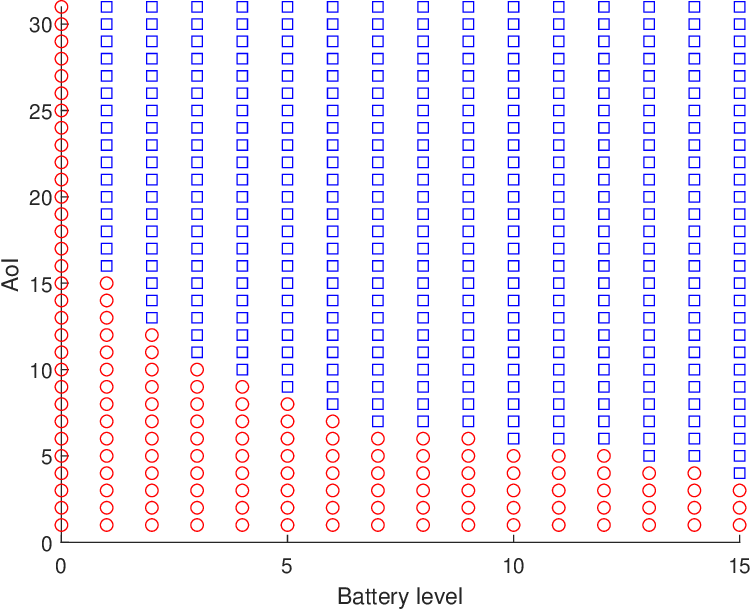}%
\label{fig3_lambda}%
}\qquad
\subfloat[$\lambda_1 = 1.0$]{%
\includegraphics[width=\fw \columnwidth]{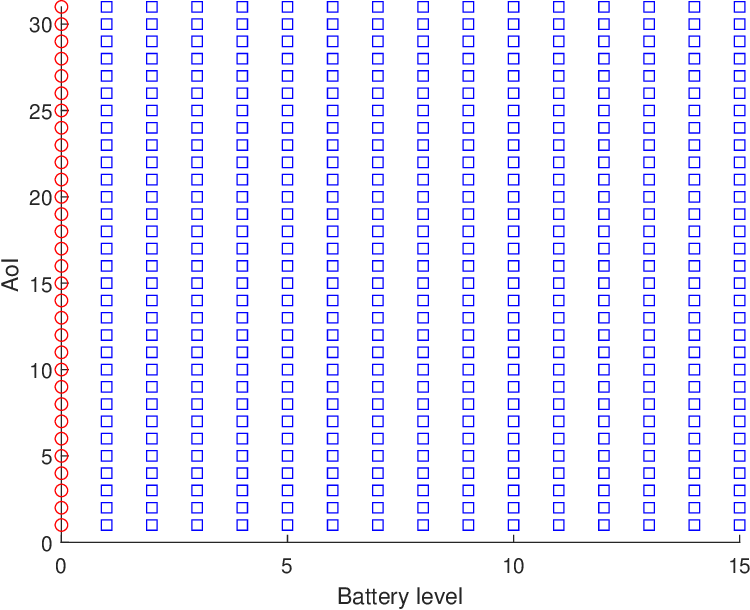}%
\label{fig4_lambda}%
}\vspace{-3mm}
\caption{Structure of an optimal deterministic policy $\pi^*_1$ obtained by {the} VIA for each state $s = \{b, {\Delta}\}$ with the transmit success probability $\xi_1 = 0.9$ for different values of the EH probability $\lambda_1$. Red circle: no command $a=0$; blue square: command $a=1$. 
}\vspace{-7mm}
\label{fig_lambda}
\end{figure*}

\begin{figure*}[t]
\centering
\subfloat [$\xi_1 = 0$]{%
\includegraphics[width=\fw \columnwidth]{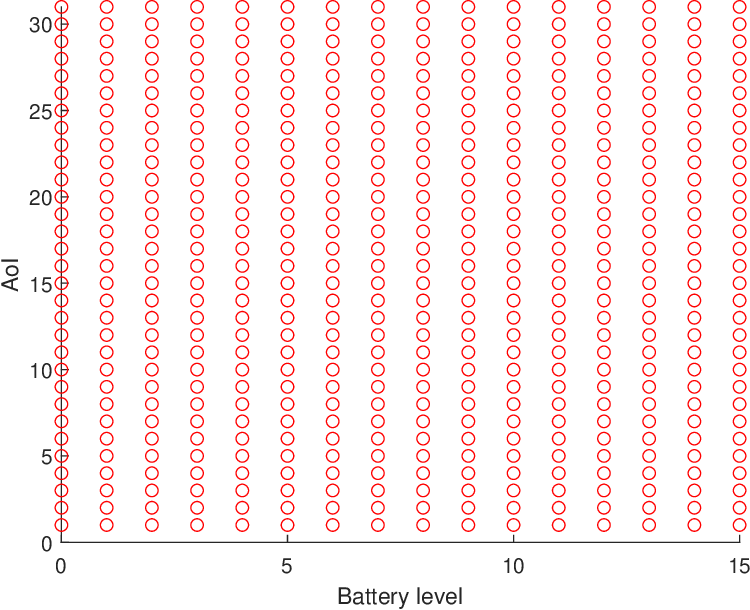}%
\label{fig1_xi}%
}\qquad
\subfloat [$\xi_1 = 0.5$]{%
\includegraphics[width=\fw \columnwidth]{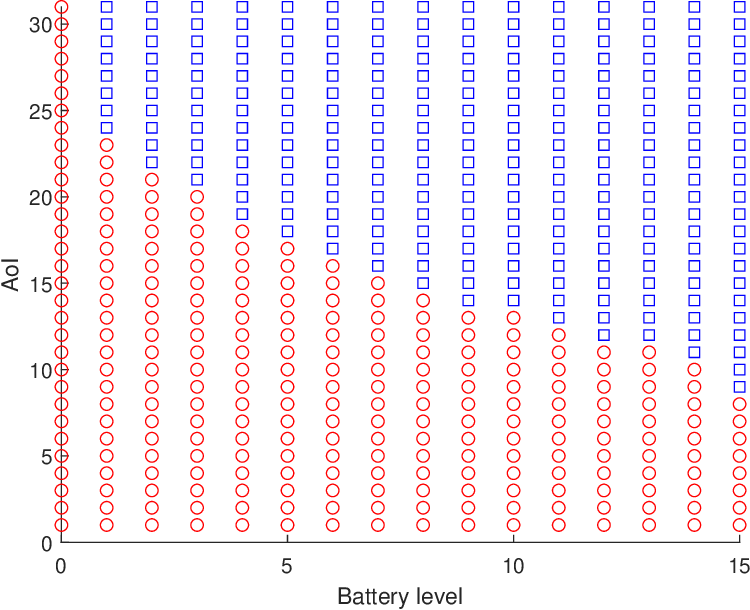}%
\label{fig2_xi}%
}\\
\subfloat[$\xi_1 = 0.7$]{%
\includegraphics[width=\fw \columnwidth]{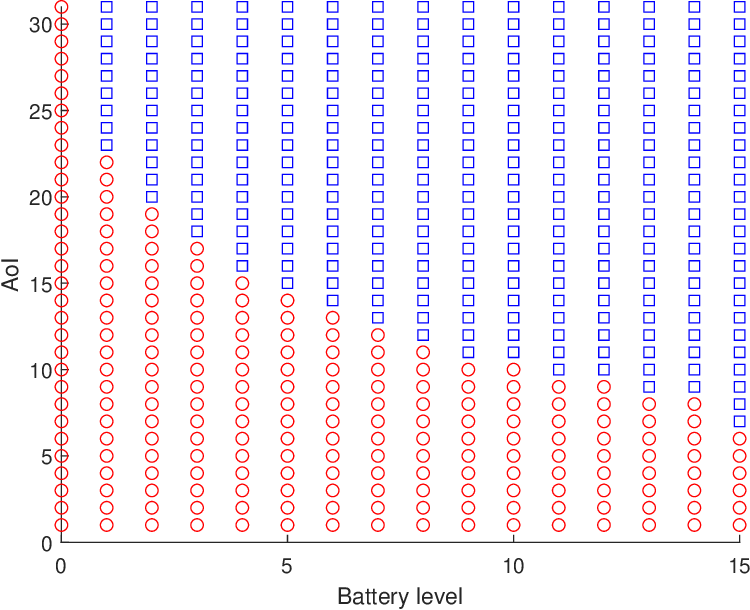}%
\label{fig3_xi}%
}\qquad
\subfloat[$\xi_1 = 1.0$]{%
\includegraphics[width=\fw \columnwidth]{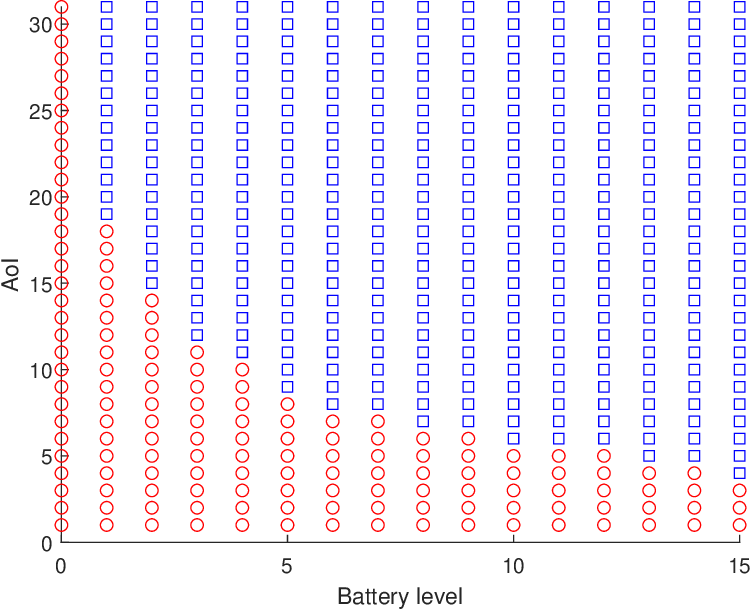}%
\label{fig4_xi}%
}\vspace{-2mm}
\caption{Structure of an optimal deterministic policy $\pi^*_1$ obtained by the VIA for each state $s = \{b, {\Delta}\}$ with the {EH} probability $\lambda_1 = 0.04$ for different values of the transmit success probability $\xi_1$. Red circle: no command $a=0$; blue square: command $a=1$.}
\label{fig_xi}\vspace{-7mm}
\end{figure*}

Fig.~\ref{fig_xi} illustrates the threshold-based structure of the {obtained} optimal deterministic policy for different values of the transmit success probability $\xi_1$ with the {EH} probability $\lambda_1 = 0.04$. 
Figs.~\ref{fig_xi}(a)--(d) illustrate that the command region expands by increasing the transmit success probability $\xi_1$. This is due to the fact that by increasing $\xi_1$, the communication link from the sensor to the edge node becomes more reliable, and thus, the edge node commands the sensor more often as it has more confidence about receiving the transmitted status update packet.
Fig.~\ref{fig_xi}(a) depicts an extreme case with $\xi_1 = 0$, in which the link from the sensor to the edge node is always in the failed state and the edge node never receives any commanded status update; to conserve the sensor's battery, the optimal action is clearly $a = 0$.

\newcommand\ftt{Xi}

\subsection{Performance and Learning Behaviour of the {Proposed} Algorithms}

\newcommand\fwp{.42}
\begin{figure*}[t]
\centering

\subfloat [Average cost for sensor 1, $\bar{C}_1$, with $\lambda_1 = 0.04$ and $\xi_1 = 0.15$]{%
\includegraphics[width=\fwp \columnwidth]{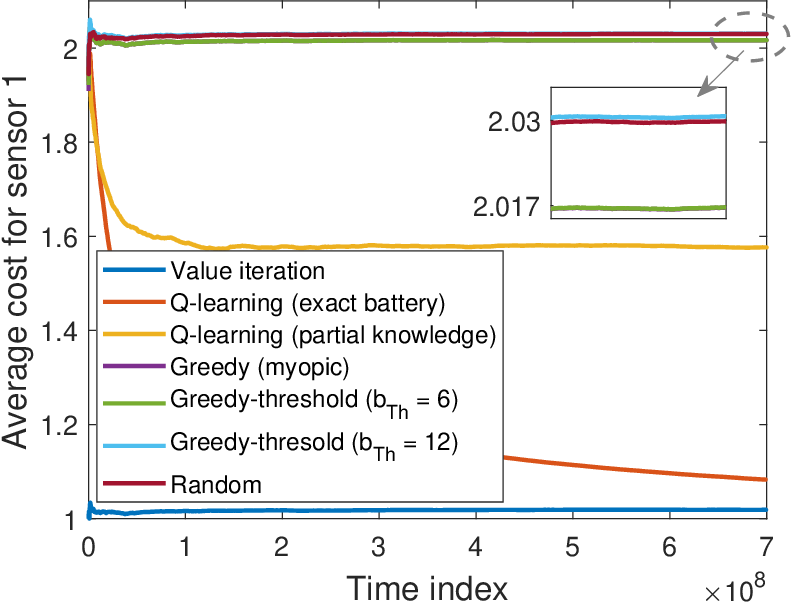}%
\label{fig1_learning}%
}\qquad
\subfloat [Average cost for sensor 2, $\bar{C}_2$, with $\lambda_2 = 0.05$ and $\xi_2 = 0.15$]{%
\includegraphics[width=\fwp \columnwidth]{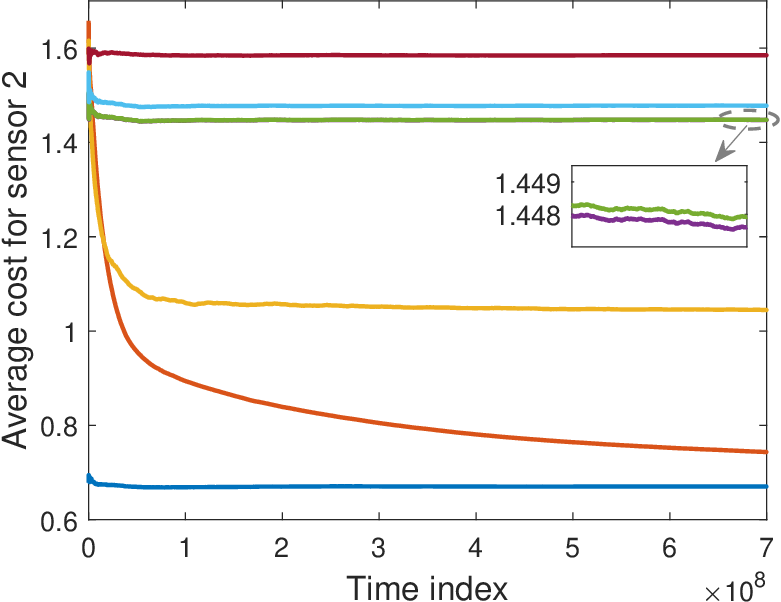}%
\label{fig2_learning}%
}\\
\subfloat[Average cost for sensor 3, $\bar{C}_3$, with $\lambda_3 = 0.06$  and $\xi_3 = 0.15$]{%
\includegraphics[width=\fwp \columnwidth]{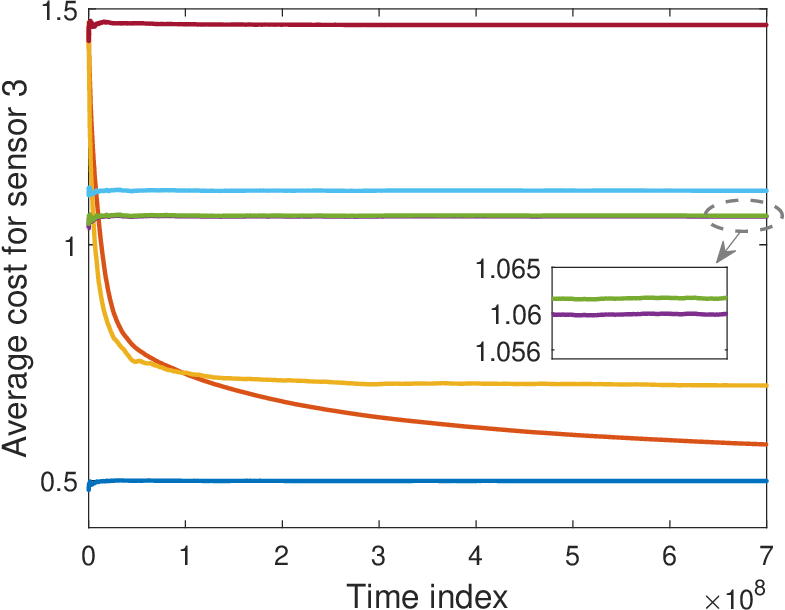}%
\label{fig3_learning}%
}\qquad
\subfloat[Average cost for all the sensors $\bar{C}$]{%
\includegraphics[width=\fwp \columnwidth]{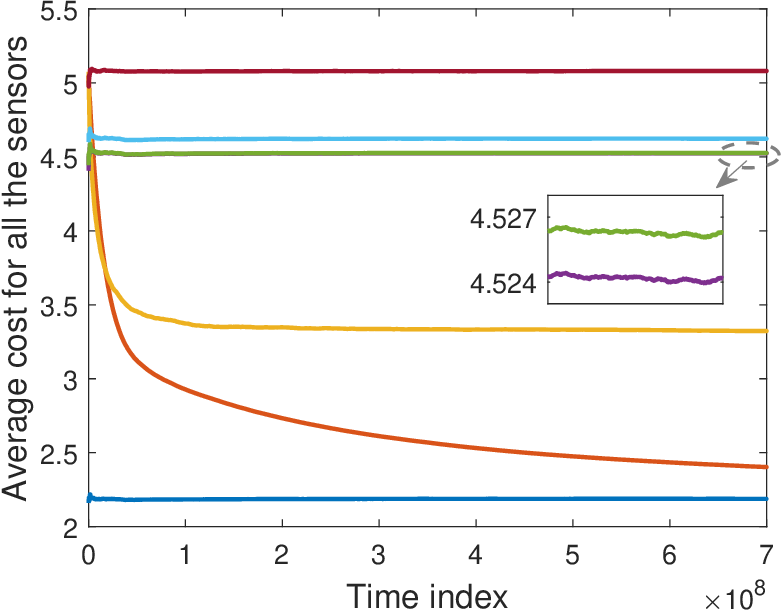}%
\label{fig4_learning}%
}\vspace{-2mm}
\caption{Learning behaviour of the proposed {VIA} and Q-learning algorithms in comparison to baseline policies.}\vspace{-7mm}
\label{fig_learning}
\end{figure*}

We investigate the performance and learning  behaviour of the proposed Q-learning algorithms with exact and partial battery knowledge.
To this end, we analyze the performance of the proposed algorithms in terms of the long-term average costs defined in \eqref{average_cost} and \eqref{average_cost_persensor}. 
{As a remark, the {VIA} serves as a lower bound to the proposed Q-learning algorithms since it knows the exact statistical model of the environment, and consequently, the state transition probabilities of the underlying MDP. Similarly, the Q-learning method with the exact battery knowledge {(referred to as \textit{Q-learning-exact} hereinafter)} is a lower bound to the Q-learning algorithm having only the partial battery knowledge {(referred to as \textit{Q-learning-partial} hereinafter)}.}

For comparison, we consider two baseline policies: \textit{greedy (myopic)}, {\textit{greedy-threshold}}, and \textit{random} policy.
In the greedy policy, whenever the value of physical quantity $f_k$ is requested (i.e., $r_k(t)=1$), the edge node commands sensor $k$ (i.e., $a_k(t)=1$), {regardless of the battery stage and AoI}; sensor $k$ sends a status update if the battery is non-empty, i.e., $b_k(t) \geq 1$.
{In the greedy-threshold policy, whenever the value of physical quantity $f_k$ is requested (i.e., $r_k(t)=1$), the edge node commands sensor $k$ if the battery level of sensor $k$ is above a threshold $\textrm{b}_{\textrm{Th}}$ (i.e., $b_k(t) \geq \textrm{b}_{\textrm{Th}}$).
Note that the greedy-threshold policy with $\textrm{b}_{\textrm{Th}} = 1$ is equivalent to the greedy (myopic) policy.}
In the random policy, whenever the value of physical quantity $f_k$ is requested (i.e., $r_k(t)=1$), the edge node selects a random action  $a_k(t)\in \{0,1\}$ according to the discrete uniform distribution.

Fig.~\ref{fig_learning} depicts the performance of each algorithm {for the {EH} probabilities $\lambda_1=0.04$, $\lambda_2=0.05$, and $\lambda_3=0.06$, and the transmit success probabilities $\xi_k=0.15$, $\forall{k\in\mathcal{K}}$.} Figs.~\ref{fig_learning}(a)--(c) are associated with the per-sensor long-term average cost ($\bar{C}_k$) for sensor 1, 2, and 3, respectively.  Fig.~\ref{fig_learning}(d) illustrates the long-term average cost over all the sensors ($\bar{C}$).



As it is shown in Fig.~\ref{fig_learning}(d), Q-learning-exact performs close to the {VIA} and the proposed RL algorithms outperform the baseline methods in terms of the long-term average cost.
{The figures show that among the greedy-threshold baseline policies, the greedy (myopic) policy ($\textrm{b}_{\textrm{Th}}=1$) results in the best performance.} 
Q-learning-exact, and also the {VIA}, reduce the average cost approximately
by a factor of 2 compared to the greedy algorithm. Furthermore, the average cost decreases roughly $30~\%$ for Q-learning-partial
compared to the (myopic) greedy algorithm.

{Interestingly, the gap between Q-learning-partial and Q-learning-exact is small, when the EH probability is high enough.} As it is shown in Figs.~\ref{fig_learning}(a)--(c), the largest gap occurs for the sensor with the lowest EH probability, i.e., sensor 1; on the contrary, the smallest gap is obtained for sensor 3 having the highest EH probability. 
This is due to the fact that when the energy becomes scarce, the edge node receives status updates more rarely; consequently, the information about the battery levels at the edge node becomes more outdated, i.e., more uncertain, inhibiting the capability of Q-learning-partial to take near-optimal actions as taken by Q-learning-exact. 
Overall, Fig.~\ref{fig_learning} demonstrates that the proposed algorithm for a realistic scenario has high performance even if the edge node performs actions based on the outdated battery information.

{In Fig.~\ref{fig_learning}(a), the greedy policy performs as poorly as the random policy, because the EH probability is low, and thus, it is highly sub-optimal to command the sensor at all states.}
{As it can be seen in Figs.~\ref{fig_learning}(a)--(c),}
the lowest long-term average cost is associated with the sensor that has the highest {EH} probability, i.e., sensor 3. This is because sensor 3 harvests energy more often, and thus, it can send status updates more frequently upon receiving a command from the edge node. Recall that the command region enlarges by increasing the  EH probability, i.e., the edge node commands the sensor more frequently.




{By comparing Figs.~\ref{fig_learning}(a)--(c) with each other, we observe that by increasing the { EH} probability $\lambda_k$ the  long-term average cost for the {VIA}, and also for the Q-learning,  moves toward the long-term average cost for the greedy policy. This is because by increasing the {EH} probability, the command region enlarges, and thus, an optimal policy tends to the greedy policy.}



\newcommand\fwavg{.47}
\begin{figure*}[t]
\centering 
\subfloat [Number of sensors $K = 4$]{%
\includegraphics[width=\fwavg \columnwidth]{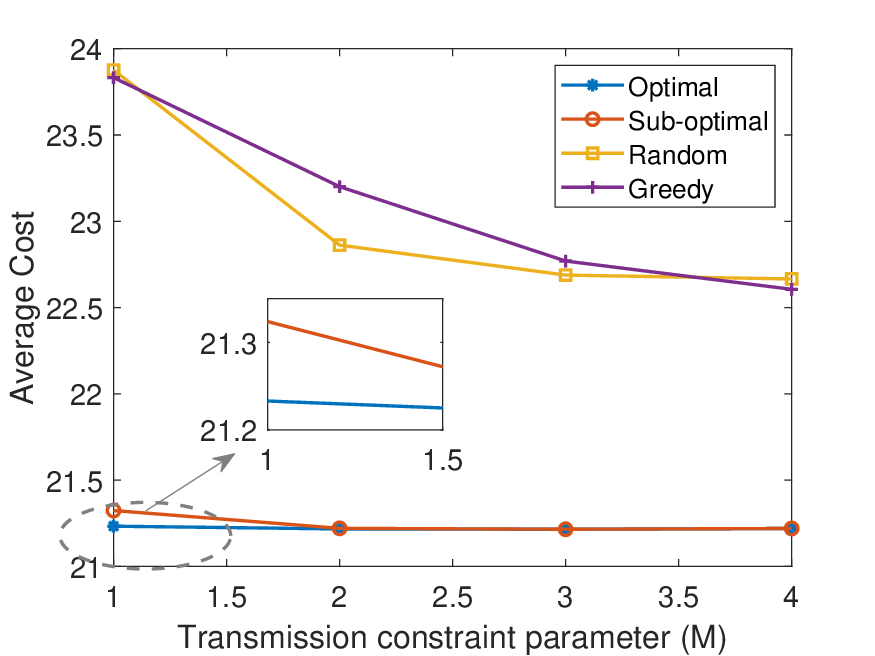}%
\label{fig1_avg}%
}\qquad
\subfloat [Number of sensors $K = 25$]{%
\includegraphics[width=\fwavg \columnwidth]{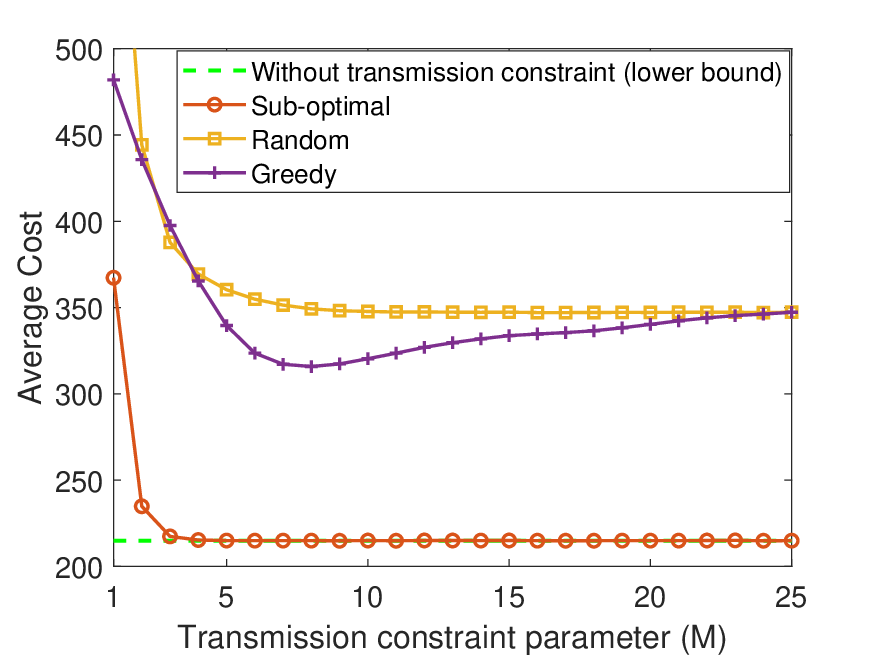}%
\label{fig2_avg}%
}\vspace{-2mm}
\caption{Performance of the proposed optimal and sub-optimal policies under the transmission limitation in comparison to the baseline policies.}\vspace{-7mm}
\label{fig_avg}
\end{figure*}

\subsection{Performance under the Transmission Constraint}\label{subsec:sim_coupled}
We investigate the performance of the proposed optimal and sub-optimal solutions presented in Section~\ref{sec_limited_bw}. 
The results are obtained by averaging each algorithm over $200$ episodes whereas each episode takes $10^6$ slots.  
We compare the proposed policy with the greedy and random policies. In the greedy policy, due to transmission constraint, the edge node commands no more than $M$ sensors with the largest AoI from the set $\mathcal{W}(t) = \{r_k(t) = 1, k\in\mathcal{K}\}$ (i.e., the set of sensors whose measurements are requested by user(s)).
In Fig.~\ref{fig_avg}(a), the performance of the optimal and sub-optimal policies are compared for different values of the transmission constraint parameter $M$ in a simple scenario with $K = 4$, $B_k = 4$, $\Delta_{k,\textrm{max}} = 8$, and $p_k = 1$.
As shown, the gap between the proposed optimal and sub-optimal solutions is small, even though the complexity of the sub-optimal is significantly lower than that of the optimal solution, as discussed in Section~\ref{sec_limited_bw}. 
In Fig.~\ref{fig_avg}(b), a more realistic scenario is considered in which $K = 25$, $B_k = 7$, $\Delta_{k,\textrm{max}} = 64$, $p_k = 1$. Note that running our algorithm to find an optimal policy in this scenario is not tractable because  the state space dimension is $|\mathcal{S}| \approx 5\times10^{67}$. For the benchmarking, we also
plot the optimal policy for the case without any transmission constraint to serve as a lower bound.
As shown, the performance of  sub-optimal policy is close to the lower bound for $M \geq 2$, which shows the effectiveness of the proposed sub-optimal solution.
Furthermore, the sub-optimal policy yields roughly $50~\%$ lower average cost than the baseline methods for (almost) all values of $M$.


\section{Conclusions and Future Work}\label{sec_conclusions}
We investigated a status update control problem  in an IoT sensing network consisting of multiple users, multiple EH sensors, and a wireless edge node. We modeled the problem as an MDP and proposed two classes of RL based algorithms: a model-based VIA relying on dynamic programming, and a model-free Q-learning method. 
Furthermore, we developed a Q-learning method for the realistic case in which the edge node does not know the exact battery levels. The proposed Q-learning schemes do not need any information about the EH model. 
{We also proposed an optimal and a low-complexity sub-optimal algorithm for a massive IoT scenario where the edge node can command only a limited number of sensors.}
{Simulation results showed that an optimal policy has a threshold-based structure and the proposed RL algorithms significantly reduce the long-term average cost compared to several baseline methods.}

{Interesting future direction of this work would be to investigate the case where the edge node cannot serve the requests from all the users at one time slot, and {study} the impact of user scheduling on the age-optimal policies for the large-scale EH IoT networks. Another future direction could be to search for {optimal and/or low-complexity} algorithms under both the partial battery knowledge at the edge node and the transmission limitation.}


\section*{Appendix}
\subsection{Proof of Proposition \ref{prop1}}\label{sec_appendix_prop1}
\begin{proof}
As discussed in Section~\ref{sec_value_alg}, the optimal state-value function $v_k^*(s)$ can be computed iteratively by the VIA. In the {VIA}, the optimal state-value function of state $s$ at iteration ${n = 1,2,\ldots}$, denoted by $v^*_k(s)^{(n)}$, is updated as (see \eqref{bellman_opt_v})
\begin{equation}\label{eq_v_itr}
\begin{array}{ll}
v^*_k(s)^{(n)} &=\min_{a\in\mathcal{A}_k} \sum_{s' \in \mathcal{S}_k} \mathcal{P}_k (s'|s,a) \left[ c_k(s,a) + \gamma v^*_k(s')^{(n-1)} \right] \\& = \min_{a \in \mathcal{A}_k} q^*_k(s,a)^{(n-1)} ,~\forall s \in \mathcal{S}_k.
\end{array}
\end{equation}
Thus, an optimal policy at $n$th iteration is given by $\pi^*_k(a|s)^{(n)} = \mathds{1}_{\{a = \argmin_{a \in \mathcal{A}_k} q^*_k(s,a)^{(n)}\}}.$
Accordingly, an optimal action in state $s$ at $n$th iteration, denoted by $a_k^*(s)^{(n)}$, {reads} as
\begin{equation}\label{eq_action_itr}
 a_k^*(s)^{(n)} =  \argmin_{a \in \mathcal{A}_k} q^*_k(s,a)^{(n)}.
\end{equation}
For any arbitrary initialization $v_k^*(s)^{(0)}$, the sequence $\{v_k^*(s)^{(n)}\}$ can be shown to converge to the optimal state-value function $v_k^*(s)$  \cite[Sect.~4.4]{sutton2018reinforcement}, i.e.,
\begin{equation}\label{eq_converge_v}
   \lim_{n\rightarrow \infty} v_k^*(s)^{(n)} = v_k^*(s).
\end{equation}

(i) In order to prove that $v_k^*(s)$ is non-decreasing with respect to the AoI, we define two states $s = \{b,\Delta \}$ and $\sbar = \{b,\Deltabar \}$, where $\Deltabar \geq \Delta$, and show that $v_k^*(\sbar) \geq v_k^*(s)$. According to \eqref{eq_converge_v}, it suffices to prove that $v_k^*(\sbar)^{(n)} \geq v_k^*(s)^{(n)}$, $\forall{n}$. We prove this by mathematical induction. 
{The  initial values can be chosen arbitrarily, e.g., $v_k^*(s)^{(0)} = 0$ and $v_k^*(\sbar)^{(0)} = 0$, thus, the relation $v_k^*(\sbar)^{(n)} \geq v_k^*(s)^{(n)}$ holds for $n = 0$.} Assume that
$v_k^*(\sbar)^{(n)} \geq v_k^*(s)^{(n)}$ for some $n$. We need to prove that $v_k^*(\sbar)^{(n+1)} \geq v_k^*(s)^{(n+1)}$ as well. From \eqref{eq_v_itr} and \eqref{eq_action_itr}, we have 
\begin{equation}
\begin{array}{ll}
v^*_k(s)^{(n+1)} - v^*_k(\sbar)^{(n+1)} & = \min_{a \in \mathcal{A}_k} q^*_k(s,a)^{(n)} -  \min_{a \in \mathcal{A}_k} q^*_k(\sbar,a)^{(n)} \\ &=   q^*_k\big(s,a_k^*(s)^{(n)} \big)^{(n)} -  q^*_k\big(\sbar,a_k^*(\sbar)^{(n)} \big)^{(n)}\\ &\overset{(a)}{\leq} q^*_k\big(s,a_k^*(\sbar)^{(n)} \big)^{(n)} -  q^*_k\big(\sbar,a_k^*(\sbar)^{(n)}\big)^{(n)},
\end{array}
\end{equation}
where (a) follows from the fact that taking action $a_k^*(\sbar)^{(n)}$ in state $s$ is not necessarily optimal. We show that  ${q^*_k\big(s,a_k^*(\sbar)^{(n)}\big)^{(n)} -  q^*_k\big(\sbar,a_k^*(\sbar)^{(n)}\big)^{(n)} \leq 0}$ for all possible actions ${a_k^*(\sbar)^{(n)} \in \{0,1\}}$. We present the proof for the case corresponding to \eqref{transition_case4} where $b \geq 1$ and $a_k^*(\sbar)^{(n)} = 1$; for the other three cases \eqref{transition_case1}--\eqref{transition_case3}, the proof follows similarly. We have
\begin{equation}
    \begin{array}{ll}
    &q^*_k(s,1)^{(n)} -  q^*_k(\sbar,1)^{(n)}\notag \\ &=  \sum_{s' \in \mathcal{S}_k} \mathcal{P}_k (s'|s,1) \left[ c_k(s,1) + \gamma v^*_k(s')^{(n)} \right] \notag - \sum_{\sbar^\prime \in \mathcal{S}_k} \mathcal{P}_k (\sbar^\prime|\sbar,1) \left[ c_k(\sbar,1) + \gamma v^*_k(\sbar^\prime)^{(n)} \right]\notag
    \\& \overset{(a)}{=} \lambda_k \xi_k \big(1 + \gamma v^*_k(b,1)^{(n)} \big) + (1-\lambda_k) \xi_k \big(1 + \gamma v^*_k(b-1,1)^{(n)} \big) \notag 
    \\ & + \lambda_k(1-\xi_k)\big(\min\{\Delta+1,\Delta_{k,\text{max}}\} + \gamma v^*_k(b,\min\{\Delta+1,\Delta_{k,\text{max}}\})^{(n)} \big)\notag
    \\ & + (1-\lambda_k) (1-\xi_k)\big(\min\{\Delta+1,\Delta_{k,\text{max}}\} + \gamma v^*_k(b-1,\min\{\Delta+1,\Delta_{k,\text{max}}\})^{(n)} \big)\notag
    \\ &- \lambda_k \xi_k \big(1 + \gamma v^*_k(b,1)^{(n)} \big) - (1-\lambda_k) \xi_k \big(1 + \gamma v^*_k(b-1,1)^{(n)} \big) \notag \\ & - \lambda_k (1-\xi_k)\big(\min\{\Deltabar+1,\Delta_{k,\text{max}}\} + \gamma v^*_k(b,\min\{\Deltabar+1,\Delta_{k,\text{max}}\})^{(n)} \big)\notag
    \\ & - (1-\lambda_k) (1-\xi_k)\big(\min\{\Deltabar+1,\Delta_{k,\text{max}}\} + \gamma v^*_k(b-1,\min\{\Deltabar+1,\Delta_{k,\text{max}}\})^{(n)} \big)\notag
    \end{array}
\end{equation}
\begin{equation}
    \begin{array}{ll}
     & = (1-\xi_k) \underbrace{\big(\min\{\Delta+1,\Delta_{k,\text{max}}\}  - \min\{\Deltabar+1,\Delta_{k,\text{max}}\} \big)}_{(b)\leq 0}\notag \\ & + \gamma \lambda_k (1-\xi_k)  \underbrace{\big( v^*_k(b,\min\{\Delta+1,\Delta_{k,\text{max}}\})^{(n)} - v^*_k(b,\min\{\Deltabar+1,\Delta_{k,\text{max}}\})^{(n)} \big)}_{(c)\leq 0} \notag
    \\ & + \gamma (1-\lambda_k) (1-\xi_k) \underbrace{ \big( v^*_k(b-1,\min\{\Delta+1,\Delta_{k,\text{max}}\})^{(n)} - v^*_k(b-1,\min\{\Deltabar+1,\Delta_{k,\text{max}}\})^{(n)} \big)}_{(d)\leq 0} \leq 0,\notag
    \end{array}
\end{equation}
where in step (a) we use the result of \eqref{transition_case4}, step (b) follows from the assumption $\Delta \leq \Deltabar$, and steps (c) and (d) follow from the induction assumption.

(ii) In order to prove that $v_k^*(s)$ is non-increasing with respect to the battery level, we define two states $s = \{b,\Delta \}$ and $\sbar = \{\bbar,\Delta \}$, where $\bbar \geq b$. By using induction and following the similar steps as we have done in (i),
one can easily show that $v_k^*(s) \geq v_k^*(\sbar)$.
\end{proof}

\subsection{Proof of Proposition~\ref{prop2}}

\begin{proof}
We define states $s = \{b,\Delta \}$ and $\sbar= \{b,\Deltabar\}$, where $\Deltabar \geq \Delta$. We show that ${\delta q_k^*(s) \geq \delta q_k^*(\sbar)}$, which can be rewritten as $q_k^*(s,1) - q_k^*(\sbar,1) - q_k^*(s,0) + q_k^*(\sbar,0) \geq 0$. We present the proof for the case where $1 \leq b < B_k$; {for the other two cases, i.e., $b = 0$ and $b = B_k$, the proof follows similarly. We have}
\begin{equation}
    \begin{array}{ll}
    & q_k^*(s,1) - q_k^*(\sbar,1) - q_k^*(s,0) + q_k^*(\sbar,0)\notag \\& = \sum_{s' \in \mathcal{S}_k} \mathcal{P}_k (s'|s,1) \left[ c_k(s,1) + \gamma v^*_k(s') \right] \notag - \sum_{\sbar^\prime \in \mathcal{S}_k} \mathcal{P}_k (\sbar^\prime|\sbar,1) \left[ c_k(\sbar,1) + \gamma v^*_k(\sbar^\prime) \right]\notag \\&-\sum_{s' \in \mathcal{S}_k} \mathcal{P}_k (s'|s,0) \left[ c_k(s,0) + \gamma v^*_k(s') \right] \notag + \sum_{\sbar^\prime \in \mathcal{S}_k} \mathcal{P}_k (\sbar^\prime|\sbar,0) \left[ c_k(\sbar,0) + \gamma v^*_k(\sbar^\prime) \right]\notag
    \\& {= \lambda_k\big( 1 + \gamma v^*_k(b,1) \big) + (1-\lambda_k)\big( 1 + \gamma v^*_k(b-1,1) \big)} \notag 
    \\ & {- \lambda_k\big( 1 + \gamma v^*_k(b,1) \big) {-} (1-\lambda_k)\big( 1 + \gamma v^*_k(b-1,1) \big)} \notag \\
    & {- \lambda_k \big(\min\{\Delta+1,\Delta_{k,\text{max}}\} + \gamma v^*_k(b+1,\min\{\Delta+1,\Delta_{k,\text{max}}\}) \big)} \notag \\&{- (1-\lambda_k) \big(\min\{\Delta+1,\Delta_{k,\text{max}}\} + \gamma v^*_k(b,\min\{\Delta+1,\Delta_{k,\text{max}}\}) \big)} \notag \\ & {+ \lambda_k \big(\min\{\Deltabar+1,\Delta_{k,\text{max}}\} + \gamma v^*_k(b+1,\min\{\Deltabar+1,\Delta_{k,\text{max}}\}) \big)} \notag \\& {+ (1-\lambda_k) \big(\min\{\Deltabar+1,\Delta_{k,\text{max}}\} + \gamma v^*_k(b,\min\{\Deltabar+1,\Delta_{k,\text{max}}\}) \big)} \notag
    \\ & = \underbrace{\big(\min\{{\Deltabar}+1,\Delta_{k,\text{max}}\} - \min\{{\Delta}+1,\Delta_{k,\text{max}}\} \big)}_{(a) \geq 0}\notag
    \\& + \gamma \lambda_k  \underbrace{\big( v^*_k(b+1,\min\{{\Deltabar}+1,\Delta_{k,\text{max}}\}) - v^*_k(b+1,\min\{{\Delta}+1,\Delta_{k,\text{max}}\} \big)}_{(b) \geq 0} \notag
     \\ & + \gamma (1-\lambda_k) \underbrace{ \big( v^*_k(b,\min\{{\Deltabar}+1,\Delta_{k,\text{max}}\}) - v^*_k(b,\min\{{\Delta}+1,\Delta_{k,\text{max}}\}) \big)}_{(c) \geq 0} \geq 0,\notag
    \end{array}
\end{equation}
{where step (a) follows from  the assumption $\Delta \leq \Deltabar$, and steps (b) and (c) follow from Proposition~\ref{prop1}.}
\end{proof}



\section{Acknowledgments}
This research has been financially supported by the Infotech Oulu, the Academy of Finland (grant 323698), and Academy of Finland 6Genesis Flagship (grant 318927). The work of M. Leinonen has also been financially supported in part by the Academy of Finland (grant 319485).
M. Codreanu would like to acknowledge the support of the European Union's Horizon 2020 research and innovation programme under the Marie Sk\l{}odowska-Curie Grant Agreement No. 793402 (COMPRESS NETS).
M. Hatami would like to acknowledge the support of HPY Research Foundation and Riitta ja Jorma J. Takanen Foundation.

\bibliographystyle{IEEEtran}
\begin{spacing}{1.33} 
\bibliography{Bib/conf_short,Bib/IEEEabrv,Bib/Bibliography}
\end{spacing}
\end{document}